\documentclass[acmsmall]{acmart}

\AtBeginDocument{%
  \providecommand\BibTeX{{%
    \normalfont B\kern-0.5em{\scshape i\kern-0.25em b}\kern-0.8em\TeX}}}

\setcopyright{acmcopyright}
\copyrightyear{2018}
\acmYear{2018}
\acmDOI{XXXXXXX.XXXXXXX}

\acmJournal{JACM}
\acmVolume{37}
\acmNumber{4}
\acmArticle{111}
\acmMonth{8}

\usepackage{amsfonts}
\usepackage{algorithm}

\usepackage{algcompatible}

\usepackage{graphicx}
\usepackage{float}
\usepackage{subfigure}

\begin{document}

\title{Decentralized Collaborative Learning Framework for Next POI Recommendation}

\author{Jing Long}
\email{jing.long@uq.edu.au}
\orcid{1234-5678-9012}
\affiliation{%
 \institution{The University of Queensland}
 \city{Brisbane}
 \state{QLD}
 \postcode{4072}
 \country{Australia}}

\author{Tong Chen}
\email{tong.chen@uq.edu.au}
\affiliation{%
 \institution{The University of Queensland}
 \city{Brisbane}
 \state{QLD}
 \postcode{4072}
 \country{Australia}}

\author{Nguyen Quoc Viet Hung}
\email{henry.nguyen@griffith.edu.au}
\affiliation{%
 \institution{Griffith University}
 \city{Brisbane}
 \state{QLD}
 \postcode{4111}
 \country{Australia}}

\author{Hongzhi Yin*}
\email{h.yin1@uq.edu.au}
\affiliation{%
 \thanks{*Corresponding author}
 \institution{The University of Queensland}
 \city{Brisbane}
 \state{QLD}
 \postcode{4072}
 \country{Australia}}


\begin{abstract}
  Next Point-of-Interest (POI) recommendation has become an indispensable functionality in 
  Location-based Social Networks (LBSNs) due to its effectiveness in helping people decide the next 
  POI to visit. However, accurate recommendation requires a vast amount of historical check-in 
  data, thus threatening user privacy as the location-sensitive data needs to be handled by cloud servers. Although there have been several on-device frameworks for 
  privacy-preserving POI recommendations, they are still resource-intensive when it comes to storage and 
  computation, and show limited robustness to the high sparsity of user-POI interactions. 
  On this basis, we propose a novel \underline{d}ecentralized \underline{c}ollaborative \underline{l}earning framework for POI 
  \underline{r}ecommendation (DCLR), which allows users to train their personalized models locally in a collaborative manner. DCLR 
  significantly reduces the local models' dependence on the cloud for training, and can be used to expand arbitrary 
  centralized recommendation models. To counteract the sparsity of on-device user data when learning each local 
  model, we design two self-supervision signals to pretrain the POI representations on the server with 
   geographical and categorical correlations of POIs. To facilitate collaborative 
  learning, we innovatively propose to incorporate knowledge from either geographically or semantically similar users into each local model with 
  attentive aggregation and mutual information maximization. The collaborative learning process makes use of communications between devices while 
  requiring only minor engagement from the central server for identifying user groups, and is compatible with common privacy preservation mechanisms like 
  differential privacy. We evaluate DCLR with two real-world datasets, where the results show that DCLR outperforms state-of-the-art on-device frameworks 
  and yields competitive results compared with centralized counterparts.
\end{abstract}




\maketitle

\section{Introduction}
Nowadays, next Point-of-Interest (POI) recommendation has gained 
immense attention in e-commerce due to the rapid growth of Location-based Social 
Networks (LBSNs), such as Weeplace and Yelp. Such services generate large volumes 
of historical check-in data, which is valuable for understanding users' behavioural 
patterns and predicting their preferences on the next movement. Next POI 
recommendation has wide applications like mobility prediction, route planning, and 
location-based advertising~\cite{LongYan2017SPRE,chen2020multi}.

To facilitate personalized POI recommendations by analyzing users' check-in data, 
early models mainly focused on Markov chains \cite{RendleSteffen2012FMwl,
ZhaoShenglin2018SSLR} and matrix factorization (MF) \cite{10.1145/2623330.2623638}. 
Recently, models based on recurrent neural networks (RNNs) \cite{FengJie2018DPhm,
hochreiter1997long} and graph neural networks (GNNs) \cite{LiYang2021DCSf} have 
demonstrated advantages in capturing temporal and structural 
dependencies among POIs. To alleviate the data sparsity problem in POI recommendation, 
recent models exploit geographical and temporal information to represent 
the spatiotemporal connections between movements \cite{10.1145/3394486.3403252,zhao2020discovering}. 
However, those models ignore correlations of non-adjacent POIs and non-consecutive
check-ins. On this basis, Luo et al. \cite{LuoYingtao2021SSan} utilize self-attention 
layers to capture relative spatiotemporal information of all check-in activities 
along the sequence.

Given the rise of intensive computing resources and massive training data, current 
deep neural networks (DNNs) have achieved the
state-of-the-art performance in POI recommendation. In addition, such recommenders are typically trained 
in a centralized way. That is, all users' data and the recommendation model are 
centrally hosted, and both the training and inference of the recommender are 
performed on the powerful cloud server. In this paradigm, users' devices only act 
as a terminal for uploading new data and recommendation requests, as well as 
displaying the recommended POIs to users.

However, such a centralized POI recommendation paradigm brings three significant 
real-world issues. 
(1) \textit{High cost of resources}. To provide precise and fast recommendations, 
large-scale data is stored and processed on the server, which consumes excessive 
storage and computing resources. Consequently, it is financially and 
environmentally expensive to maintain such resource-intensive services.
(2) \textit{Privacy issues}. Despite the successful applications of centralized 
deep learning in various online services, users' privacy is increasingly prone 
to breaches, causing huge losses for both users and enterprises 
such as Facebook \cite{HullGordon2010Cgpi} and Netflix \cite{NarayananA2008RDoL}. 
In addition, POI recommendation is especially sensitive to privacy concerns as 
historical check-in data directly reveals users' physical trajectories \cite{wang2021fast}. However, 
if users choose not to share their personal data with the POI recommender, 
they will possibly face a decreased service quality as a consequence.
(3) \textit{Weak resilience}. Centralized POI recommendation highly relies on 
the stability of the server and internet connectivity. Once the server is 
impaired or overcrowded, or network quality cannot be guaranteed, the 
recommendation services are unable to ensure timeliness and may even go offline \cite{ye2022personalized,nguyen2017argument}.
Considering many tourist attractions are located in remote areas with limited 
telecom infrastructures, weakening the resilience of cloud-based POI recommendation.

As such, there has been a recent surge in developing on-device POI 
recommendation. In on-device recommender frameworks, users can keep their personal check-in 
data on their own devices, which can significantly reduce the risk of privacy 
breaches. 
Specifically, Wang et al. \cite{WangQinyong2020NPRo} deployed compressed 
models on mobile devices for secure and stable POI recommendation. In such a framework, 
all personal information including check-in data and compressed models are stored 
on users' own devices. To address the data sparsity problem, those compressed models 
inherit the knowledge from the teacher model which is trained with public data. 
However, both the teacher and student models are still trained on the cloud server and all 
users share the same model.
In addition, Guo et al. \cite{GuoYeting2021PPRw} proposed a federated learning 
framework for next POI recommendation. However, the framework is still resource-intensive 
for storage and computation, because the central server is responsible for 
collecting and aggregating locally trained models, as well as redistributing the 
aggregated model to all users. Moreover, all users share the same global model, which ignores the dynamics and diversity of users' spatial 
activities and interests, leading to suboptimal performance. Although some approaches \cite{wang2020group,
rao2021privacy} can cluster clients to provide group-based 
personalized models, this commonly involves the transmission of sensitive user attribute information. 
To conclude, we are still in pursuit of a decentralized POI recommendation paradigm that 
requires minimal engagements and resources from the central party, and can collaboratively 
learn personalized models amid highly sparse interaction data at the individual device level.

To this end, we propose a new decentralized POI recommendation paradigm that allows 
user devices to mutually communicate and collaborate to learn performant recommenders. 
Unlike standard federated learning, in this paradigm, the central server is only 
responsible for providing pretrained parameters (i.e., POI embeddings) and grouping similar users in a secure way, such that collaborative 
model learning is enabled for user devices within the same group.
On the server side, to alleviate the data sparsity problem and speed up the training for local models, we design two 
self-supervised pretraining tasks to inject categorical and geographical information into POI 
representations. The first task is to create correlations between POIs and their associated 
categorical tags by Mutual Information Maximization (MIM) \cite{zhou2020s3}, while the second 
one is to enhance POI representations by learning and predicting distances between 
POIs. The pretrained POI embeddings, as a key part of model parameters, will be incrementally updated with each user's on-device interaction data. 
In addition, we propose two metrics to identify similar users for collaborative 
learning. Firstly, semantic neighbors are decided by category distribution 
similarities. That is, the higher the category distribution similarity between two 
users is, the more likely they are neighbors. In addition, geographical neighbors 
are decided by the physical distances where the distance between two users is the 
minimal distance between their two sets of centroids, while multiple centroids of a 
user are obtained by clustering all visited POIs with their longitudes and latitudes.
To calculate category distribution similarity and physical distances, the server has 
to collect category distributions and centroids from devices. Although such information 
is far less sensitive compared to check-in trajectories or personalized models, only 
perturbed category distributions and centroids are uploaded to the server.

On this basis, our proposed framework is more performant than the decentralized framework
proposed by Chen et al. \cite{chen2018privacy} that also allows local models to 
exchange knowledge with geographically close users decided by the random walk theory. \cite{chen2018privacy} is subject to a performance drop compared with the state-of-the-art centralized POI recommendation 
frameworks, since the limited knowledge from a single type of neighbors 
cannot utilize the knowledge from similar users effectively. Based on two types of neighbors informed by the server, we design a novel MIM task to jointly learn and combine knowledge 
from both types of neighbors in a unified and elegant way. As such, we can obtain a 
high-quality, fully personalized local recommender for each user. 

To conclude, our contributions are listed as follows:
\begin{itemize}
\item We propose a novel \underline{d}ecentralized \underline{c}ollaborative \underline{l}earning framework for POI 
\underline{r}ecommendation (DCLR), where users receive a fully personalized local recommendation model while retaining all personal data on-device. The new 
paradigm allows for collaborative model learning with minimal dependency on the 
cloud, and is generalizable to arbitrary centralized recommendation models.

\item To alleviate data sparsity when learning each local POI recommender, we design 
two self-supervised learning objectives to make full use of the side information 
on geographical correlations and categorical tags of POIs. 

\item  We propose two metrics to identify neighbors based on categorical similarity 
and geographical distances. In addition, we propose an effective way to learn 
knowledge from both neighbors with attentive aggregation and mutual information 
maximization.

\item	We evaluate DCLR with two real-world datasets. The results show the 
effectiveness and efficiency of our proposed model. Specifically, DCLR outperforms all 
on-device frameworks and provides competitive POI recommendation accuracy compared 
with advanced centralized models.
\end{itemize}

\begin{figure*} [ht]
	\centering 
	\includegraphics[width=5.4in,height=6.4in]{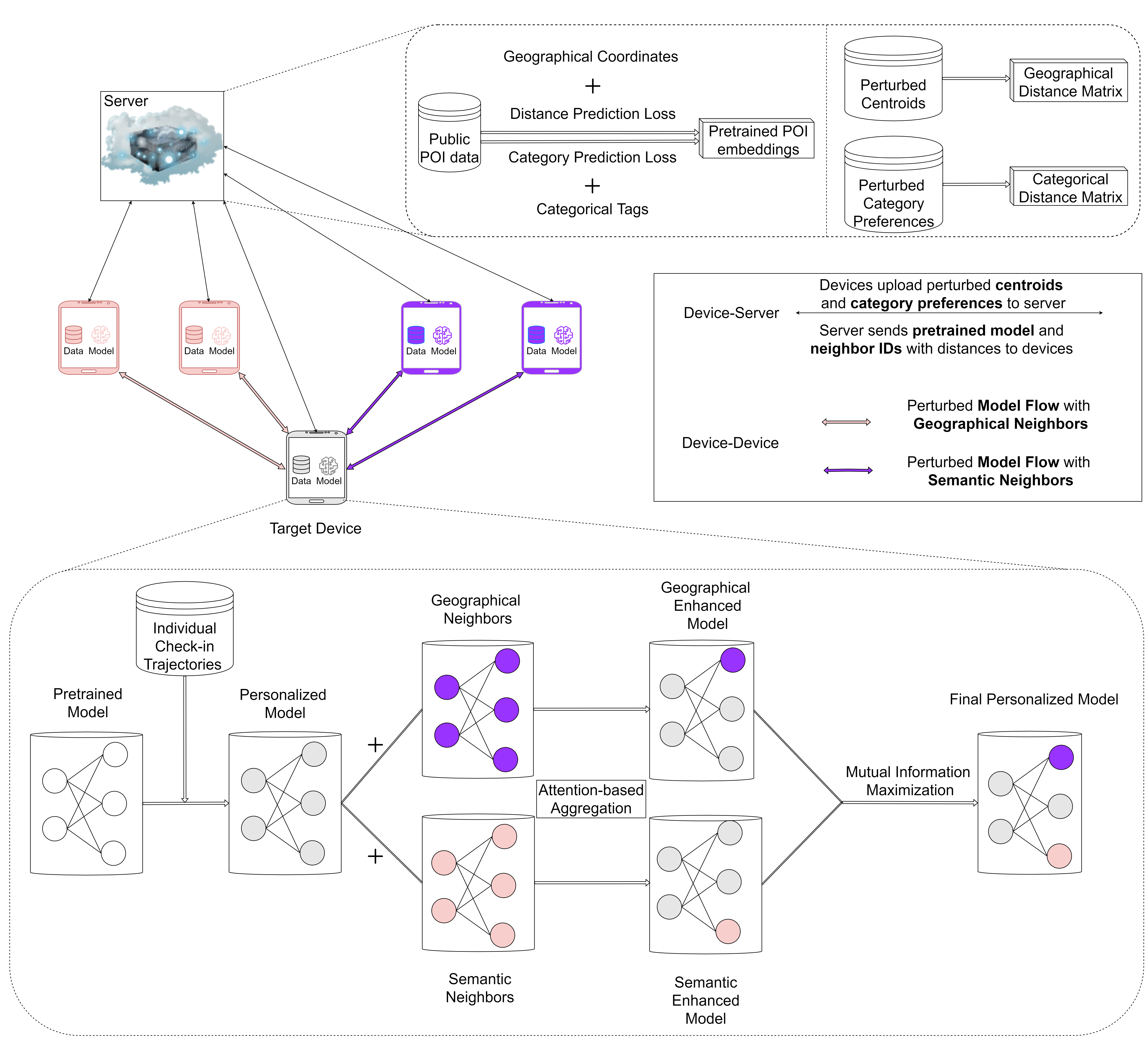}
	\caption{The overview of our proposed DCLR. All individual check-in trajectories are stored on the device side. 
  The server is responsible for providing model parameters pretrained with public data (i.e., geographical 
  coordinates and categorical tags of POIs) and identifying neighbors for each client in a privacy-preserving 
  manner. After receiving the pretrained model and neighbor IDs, each device firstly optimizes the personalized 
  model with individual check-in data, and further enhances the model with the knowledge from two types of 
  neighbors.}
	\label{fig:diagram} %% label for entire figure 
\end{figure*}

\section{PRELIMINARIES}
In this section, we first introduce important notations used in this paper, and 
then formulate our major tasks.

We denote the set of user, POI and category as
$U=\{u_1,u_2,...u_{|U|}\}$, 
$P=\{p_1,p_2,...p_{|P|}\}$, 
$C=\{c_1,c_2,...c_{|C|}\}$ 
respectively. Each POI $p\in P$ is associated with a category $c\in C$ 
and a geographical coordinates $(lon_p,lat_p)$.

\textbf{Definition 1: Check-in Activity}. A check-in activity of a user indicates a user $u\in U$ has visited a POI $p\in P$ at a timestamp 
$t$. It can be denoted by a tuple $x=(u,p,t)$.

\textbf{Definition 2: Check-in sequence}. A check-in sequence contains $M$ consecutive 
check-in activities visited by a user $u_i$, denoted by $X(u_i)=\{x_1,x_2,...,x_M\}$.

\textbf{Task 1: Decentralized Next POI Recommendation}. In our decentralized framework, 
the server is only responsible for providing pretrained parameters (i.e., POI embeddings) and identifying 
neighbors for each client in a privacy-preserving manner. 
No user profiles are used by either local models or the server. After receiving the 
anonymized neighbor IDs from the server, each mobile device will obtain and aggregate knowledge from its 
neighbors to update the local model. Note that each device hosts only one user's data and model. Given $X(u_i)$ and neighbor IDs, the local model is trained 
to provide a ranked list of POIs based on each user's recent interest as recommendations.

\section{THE PROPOSED FRAMEWORK}
% \begin{figure*} [htbp]
% 	\centering 
% 	\includegraphics[width=5.4in]{test1.png}
% 	\caption{Framework Diagram}
% 	\label{fig:something} %% label for entire figure 
% \end{figure*}

In this section, we formally introduce our techniques to 
build DCLR for next POI recommendation. The overview of DCLR is illustrated in Figure \ref{fig:diagram}. 
Overall, our proposed DCLR consists of: 
(1) A \textbf{base recommender} that will be deployed on each user device. 
(2) Two \textbf{self-supervised learning tasks} that enhance POI representation learning with public data on the server side. 
(3) A \textbf{neighbor identification} strategy deployed on the server that consists of two metrics for securely identifying 
similar users in collaborative learning. 
(4) A \textbf{neighbor communication} strategy to jointly learn and combine knowledge from 
both types of neighbors in an effective, secure, and on-device manner.
In what follows, we unfold the design of each module of our proposed approach.

\subsection{Base POI Recommender}
The key component of our decentralized framework is to exchange knowledge with homogeneous 
models learned with different users' own data. Thus, our 
decentralized framework is scalable and flexible, which can be applied to almost any next POI recommenders. 
Our work is based on Spatio-Temporal Attention 
Network (STAN) \cite{LuoYingtao2021SSan}, which is one of the newest POI recommender models. 
As our main contributions lie in the decentralized collaborative learning paradigm rather than the base model, we keep the introduction of STAN brief.

We use $\textbf{e}_p\in\mathbb{R}^d$ and $\textbf{e}_t\in\mathbb{R}^d$ to denote the embedded 
representations of POI, and time, respectively. For time discretization, we follow \cite{LuoYingtao2021SSan} to divide continuous 
timestamps into $7\times 24 = 168$ intervals to represent the exact hours in a week. 
Therefore, the input dimensions of the embeddings $\textbf{e}_p$ and $\textbf{e}_t$ are $|P|$ 
and 168, and the output for each check-in activity $x$ 
is $\textbf{e}_x=\textbf{e}_p+\textbf{e}_t\in\mathbb{R}^d$.
Then, the input embedding of each user's sequence is $\textbf{X}_u=[\mathbf{e}_{x_1},\mathbf{e}_{x_2}
,...,\textbf{e}_{x_M}]\in\mathbb{R}^{M\times d}$.
STAN further encodes the spatiotemporal gaps between two check-ins $x_a$ and $x_b$ via 
$\textbf{e}_{ab}^{\Delta}\in\mathbb{R}^d$: 
\begin{equation}
  \textbf{e}_{ab}^{\Delta}=\Delta_{ab}^s \times \textbf{e}_{\Delta_s} + 
  \Delta_{ab}^t \times \textbf{e}_{\Delta_t} 
\end{equation}
\noindent{where $\textbf{e}_{\Delta_s}$ and $\textbf{e}_{\Delta_t}$ are two unit embeddings to represent a 
specific amount of spatial (e.g., one kilometer) or time (e.g., one hour) difference, $\Delta_{ab}^s$ and $\Delta_{ab}^t$are the 
true spatiotemporal differences of $x_a$ and $x_b$ (e.g., 10 kilometers and 5 hours).}
On this basis, the embedding of trajectory spatio-temporal relation matrix is 
$\Delta\in\mathbb{R}^{M\times M}$:
\begin{equation}
  \Delta= \left[ \begin{array}{cccc}
    e_{11}^{\Delta '} & e_{12}^{\Delta '} & ... & e_{1M}^{\Delta '}\\
    e_{21}^{\Delta '} & e_{22}^{\Delta '} & ... & e_{2M}^{\Delta '}\\
    ... & ... & ... & ...\\
    e_{M1}^{\Delta '} & e_{M2}^{\Delta '} & ... & e^{\Delta '}_{MM}
    \end{array} 
    \right ]
\end{equation}
\noindent{where} $e^{\Delta '}_{ab}$ is the element-wise sum of $\textbf{e}_{ab}^{\Delta}$.
Then, we adopt the self-attention mechanism to further combine embedded sequences and 
spatiotemporal differences. Given three parameters $\textbf{W}_Q$, $\textbf{W}_K$, $\textbf{W}_V \in\mathbb{R}^{d \times d}$, 
the final embedded sequence $\textbf{E}_u\in\mathbb{R}^{M \times d}$ is defined as follows:
\begin{equation}
  \textbf{E}_u = Softmax(\frac{\textbf{QK}^T+\Delta}{\sqrt{d}})\cdot \textbf{V} 
\end{equation}
\noindent{where $\textbf{Q}  = \textbf{X}_u\textbf{W}_Q$, $\textbf{K}  = \textbf{X}_u\textbf{W}_K$, $\textbf{V}  = \textbf{X}_u\textbf{W}_V$}. 
Given the final embedding of the sequence $\textbf{E}_u\in\mathbb{R}^{M \times d}$, 
the embedding of $h$ candidate POIs $\textbf{E}_{cand}=[\textbf{e}_{p_1},\textbf{e}_{p_2}
,...,\textbf{e}_{p_h}]\in\mathbb{R}^{h\times d}$, 
and the spatio-temporal relation matrix 
$\Delta_{cand}\in\mathbb{R}^{h \times M}$ for those $h$ candidate POIs and $M$ visited POIs,
the likelihood $\alpha\in\mathbb{R}^{h}$ of candidate POIs to be visited by a user is calculated as:
\begin{equation}
  \alpha = Sum(Softmax(\frac{\textbf{E}_{cand}\textbf{E}_u^T+\Delta_{cand}}{\sqrt{d}}))
\end{equation}
\noindent{where $Sum(\cdot)$ is the sum of the last dimension.} 

Afterwards, for each user's on-device model, we adopt cross-entropy to define the POI prediction loss for optimization:
\begin{equation}
  L_{POI} = -\sum\limits_{n=1}^{N_{pos}}\left(\log\sigma(\alpha_{i})+\frac{1}{N_{neg}}\sum\limits_{j=1, j\neq i}^{N_{neg}} \log(1-\sigma(\alpha_j))\right)
\end{equation}
where 
$N_{pos}$ is the number of positive samples for a user, 
$N_{neg}$ is the number of negative samples for each positive sample $p_i$ that are randomly 
drawn from users' unvisited POIs.
$\sigma(\cdot)$ is the sigmoid function, while
$\alpha_i$ and $\alpha_j$ are respectively the scores assigned to the positive and negative samples.

\subsection{Self-Supervised Tasks for POI Embedding Pretraining}
The sparsity of user-POI interactions has been a long-standing issue in POI recommendation, 
which deteriorates further in decentralized settings as each 
client now needs to independently learn a local model with only the on-device user data. 
Though parameter sharing via a central server (e.g., federated learning) 
can help alleviate this problem, the low-quality models submitted by clients at the first 
place harms the convergence efficiency and bottlenecks the performance of the final model.
On this basis, we propose two self-supervision signals to enhance POI representations with the 
geolocations and categorical tags that are widely available in POI recommendation datasets. 
The self-supervised tasks are used for pretraining the POI embeddings on the serve 
before they are further optimized on-device with users' own interactions. 
Compared with self-supervised learning tasks requiring interaction sequences, e.g., mask prediction 
tasks \cite{zhou2020s3}, the parameter pretraining in DCLR only involves the commonly available and insensitive public 
data about POIs, i.e., geolocations and category tags. In contrast, if we perform mask prediction on the server, all users 
need to upload their personal trajectories to the server, bringing significant problems including high communication costs 
and privacy issues. Mask prediction tasks are unsuitable to be performed on-device either, as POI embeddings trained only with 
the target user's trajectories are heavily biased while it is infeasible to obtain travel records from other users.

% Thanks for your suggestion. In our revised version, we further provide detailed reasons of choosing such 
% self-supervised learning tasks over other options available.
% Our proposed self-supervised learning tasks (geolocation-guided self-supervised learning and category-aware 
% self-supervised learning are responsible for pretraining POI embeddings on the server side before they are 
% distributed to user devices for finetuning. To preserve user privacy, this pretraining process only involves the 
% commonly available and insensitive public data about POIs, i.e., geolocations and category tags).
% In contrast, if we perform mask prediction on the server, all users need to upload their personal trajectories 
% to the server, bringing significant problems including large communication cost and privacy issues. Mask prediction 
% tasks are unsuitable to be performed on-device either, as POI embeddings trained only with the target user’s trajectories 
% are heavily biased while it is infeasible to obtain travel recordes from other users.

On one hand, geographical information is indispensable in next POI recommendation as users' preferences 
on next POIs are highly dependent on their distances to users' current POIs 
\cite{10.1145/3394486.3403252,liu2014exploiting}. However, current POI recommenders only capture the geographical correlation 
between observed POIs in trajectories. Thus, we further encode the geographical properties with a pairwise 
distance prediction task between POIs.
On the other hand, while users' check-ins on specific POIs are sparse, their interactions at the category level 
(e.g., restaurants, entertainment, etc.) are relatively denser. The prediction of category-wise preferences 
is able to provide strong predictive signals on user interests at the POI level. For example, a user's strong 
preference on entertainment venues indicates she/he is likely to visit POIs like cinemas and pubs under this 
category. Therefore, we create correlations between POIs and their associated categories by maximizing 
their mutual information \cite{zhou2020s3}.

\textbf{Geolocation-guided Self-supervised Learning.} 
Firstly, we enhance POI representations by injecting
geographical information. Each POI $p$ is associated with a geographical coordinate $(lon_p,lat_p)$. 
Hence, the true distance in kilometer between any pair of POIs can be calculated by adopting Haversine function \cite{Robusto1957TheCF}:
\begin{equation}
  dis(p_i,p_j) = Haversine((lon_{p_i},lat_{p_i}),(lon_{p_j},lat_{p_j}))
\end{equation}
\noindent{Then, we assign labels for different distances: small if $dis(p_i,p_j) \leq 5km$, medium if $5km < dis(p_i,p_j) \leq 10km$, and large for distances $dis(p_i,p_j) > 10km$. Our choice of cut-off points is based on the observation from \cite{chang2018content} which is generalizable across common LBSNs. On this basis, we can enhance POI representations by learning and predicting distances between POIs. Then, we formulate the distance prediction (DP) objective below:
\begin{equation}
  L_{DP}=-\sum\limits_{\forall p_i, p_j}\sum\limits_{l=1}^{L}y^{ij}_l \log(\hat{y}^{ij}_l)
\end{equation}
where $L=3$ is the number of distance labels (i.e., small, 
medium, and large), $y^{ij}_l$ is the one-hot indicator for label $l$, and $\hat{y}^{ij}_l$ is the predicted probabilities for label $l$. The $L$-dimensional probability distribution $\hat{\textbf{y}}_{ij}$ over all distance intervals are calculated as follows:
\begin{equation}
  \hat{\textbf{y}}_{ij} = Softmax(\textbf{w}_{DP}\textbf{e}_{p_i}^T \cdot \textbf{e}_{p_j})
\end{equation}

\noindent{where} $\textbf{e}_{p_i},\textbf{e}_{p_j}\in\mathbb{R}^{d}$ are embeddings of POI $p_i$ and POI $p_j$, and $\textbf{w}_{DP}\in\mathbb{R}^{3\times 1}$ are learnable parameters. 
It is worth noting that, it makes little sense to perform distance prediction for all POI pairs 
given the limited on-device computing resources and the low possibility of travelling to long-distance POIs from the current one. Hence, for each POI, training samples consist of all short-distance POIs and a certain amount (i.e., 500 in our case) of medium-distance and long-distance POIs which are randomly selected.

\textbf{Category-aware Self-supervised Learning.} 
We further propose to enrich POI embeddings with category information by leveraging Mutual Information Maximization \cite{zhou2020s3}. Mutual information (MI) 
refers to dependencies between two variables. That is, conditioned on one variable, MI estimates certain knowledge of the other via the following:
\begin{equation}
  MI(X,Y) = P(X)-P(X|Y) = P(Y)-P(Y|X)
\end{equation}
\noindent{MI} has been widely used for learning feature representations. However, directly optimizing
this objective is computationally intractable. Instead, we maximize its lower bond via the InfoNCE 
loss \cite{kong2019mutual}. Back to our scenario, we treat a POI $p$ and its associated 
category $c$ as two views of the POI. Inspired by InfoNCE, we design a category prediction (CP) loss to maximize the mutual information between these two views, which is defined below:
\begin{equation}
  L_{CP} = -\sum_{\forall p_i} \log\frac{\exp({f_{CP}(p_i, c_{p_i}))}}{\sum\limits_{n=1}^{N_{CP}} 
  \exp({f_{CP}(p_i,c_n)})}
\end{equation}
\noindent{where} $c_{p_i}$ is a POI's associated category, 
where $N_{CP}$ negative pairs $(p_i,c_n)$ are sampled from the categorical set $C\setminus c_{p_i}$, and $f_{CP}(\cdot,\cdot)$ is a function of the pairwise similarity between the POI and the category
to a single value. In DCLR, it is implemented by a bilinear network with a scalar output:
\begin{equation}
  f_{CP}(p,c)=\sigma(\textbf{e}_p^T\textbf{W}_{CP}\textbf{e}_c)
\end{equation}
\noindent{where} $\textbf{e}_p\in\mathbb{R}^{d}$ and $\textbf{e}_c\in\mathbb{R}^{d}$ are 
embeddings of POI and category, and $\textbf{W}_{CP}\in\mathbb{R}^{d\times d}$ is a learnable parameter matrix. By optimizing the loss function, we can maximize the mutual 
information between positive pairs while minimizing that between negative pairs. 

\subsection{Privacy-aware Neigbor Identification}

In DCLR, each user's personalized model is obtained by further optimizing the network parameters and pretrained POI embeddings through the recommendation loss $L_{POI}$. 
Although self-supervised learning helps to learn better POI representations despite sparsity, it is still a challenge 
to estimate a user's exact preference with a limited amount of interactions. To address the problem, 
we propose to let each user device communicate with its neighbors who are similar to the current user. However, in our proposed framework, users only have access to their own check-in data, and thus, they cannot get the information to decide their neighbors. To tackle this issue, we ask the server to perform neighbor identification for each user on the server side. Compared with federated frameworks that also cluster similar users to facilitate better model personalization \cite{wang2020group,rao2021privacy}, DCLR avoids collecting users' raw data which is highly sensitive, and instead leverages implicit user preference indicators to support accurate yet privacy-aware neighbor identification. Furthermore, after the self-supervised pretraining task, the server is only responsible for identifying neighbors for users, and is released from the hefty role of aggregating all local models throughout the training cycle which is computationally expensive. In addition, privacy can be rigorously guaranteed when such implicit information is masked by privacy mechanisms (e.g., differential privacy) before sharing. In this section, we provide details of the privacy-aware neighbor identification on the server side. 

\textbf{Geographical Neighbors}. In the scenario of POI recommendation, if two users frequently visit venues in the same area, they possess high geographical similarity 
because they might have visited the same POIs in the past, and are likely to visit the same POIs in the future. 
In addition, users from the same or nearby POIs may share similar mobility patterns, which are easily 
influenced by the environment \cite{davies1981constituents,nguyen2017retaining}. We term such neighbors as 
\textbf{Geographical Neighbors}. On this basis, we propose to identify similar 
users by measuring the distances between the centroids (i.e., averaged coordinates) of two users' visited POIs. 
The smaller the distance between two centroids, the higher affinity two users share. However, in some cases, a single centroid coordinate is unable to comprehensively represent a user's frequent 
activity zones. For example, if a user's visited POIs are distributed in two different cities, the centroid 
might be somewhere that is far away from both. In this situation, the centroid is not helpful for finding 
neighbors. To address this problem, we advocate generating multiple centroids for each user by applying k-means 
clustering \cite{macqueen1967some} on the coordinates of all her/his visited POIs. The number of centroids is 
adaptively determined for each user, such that no visited POIs have a distance above our predefined threshold (i.e., $10km$ in our case) with their 
nearest centroids. Then, we can obtain a set of centroids for each user $u_n$:
\begin{equation}
  O(u_n) = \{o^1,o^2,...,o^v\} 
\end{equation}
\noindent{where $v$ is the number of centroids. In addition, $v$ is not supposed to be the same for every user 
as different users' visited places may vary in numbers and geolocations. Afterwards, the geographical distance of two users is calculated by the function below:
\begin{equation}
  d_{geo}(u_n,u_m) = min(Haversine(o_n,o_m))
\end{equation}
\noindent{where} $\forall o_n\in O(u_n),\forall o_m\in O(u_m)$.

\textbf{Semantic Neighbors}. Meanwhile, it is worth pointing out that not only users who are physically close to each other are regarded similar, 
but also users can be highly relevant at the semantic level. Having visited similar places indicates that users 
have the same interests, and thus, such users are considered to be similar. Such neighbors are termed \textbf{Semantic Neighbors}. 
Although both categorical preferences
and POI-level preferences can reflect such affinity between users \cite{han2020contextualized,yu2019generating,yu2020enhance}, we leverage categorical preferences as the metric of semantic similarity for the following reasons. Firstly, people with the same interests might locate far away from each other and are hence unlikely to have similar POI-level preferences. Furthermore, compared with users' check-ins on specific POIs, their interactions at the category level are relatively denser, making categorical preference a more accurate indicator of semantic similarity. Lastly, revealing users' POI-level preferences  in the cyber environment are more prone to privacy breaches.

Formally, we use $CP(u_n)=\{cp_1^n,cp_2^n,...cp_{|C|}^n\}$ to denote the user's distribution over $|C|$ POI categories, which 
can be directly derived from all visited POIs in the user's training data. For example, a dataset has three categories 
including Bar, Music Venue and Steakhouse, and a user has visited Bar for 2 times, Music Venue for 3 times, and 
Steakhouse for 5 times. Then, the category distribution of the user is $CP(u_n)=\{0.2,0.3,0.5\}$. Thus, we adopt Kullback-Leibler (KL) divergence \cite{botev2010kernel,Giantomassi2015ElectricMF} to quantify the distance between two users' categorical preferences, which is formulated as follows: 
\begin{equation}
  d_{cat}(u_n,u_m) = \sum_{c=1}^{|C|}cp_c^n\cdot \log\frac{cp_c^n}{cp_c^m}
\end{equation}

Based on the methods above, the server is able to calculate geographical and categorical distances between any pair of users, and we use $\textbf{D}_{geo}\in\mathbb{R}^{U\times U}_{\geq 0}$ and $\textbf{D}_{cat}\in\mathbb{R}^{U\times U}_{\geq 0}$ to denote the generated geographical and categorical distance matrices among users. The $i$-{th} row of the two
matrices represents geographical distances and categorical distances between the $n$-th user and all other users. 
In addition, the $q$ nearest neighbors of the $n$-th user can be decided by sorting the $n$-th row, excluding the $n$-th user, and 
selecting $q$ users with the smallest distances. Then, for user $u_n$, we 
use $N_{geo}(u_n)=\{(u_m, d_{geo}(u_n, u_m))\}_{m=1}^{q}$ and $N_{cat}(u_n)=\{(u_{m'}, d_{cat}(u_n, u_{m'}))\}_{m'=1}^{q}$ to carry the identified geographical and categorical neighbor IDs and their  distances with $u_n$. Specifically, those two sets are the information that the server will handover to all users for subsequent computations after the neighbor identification.

%$N_g^i=\{\Theta_1^g,\Theta_2^g,...,\Theta_q^g\}$ and $N_c^i=\{\Theta_1^c,\Theta_2^c,...,\Theta_q^c\}$

\textbf{Incorporating Differential Privacy.} In our proposed framework, we maintain a server to collect centroids $O(u_n)$ and categorical preferences $CP(u_n)$ of all users. After calculating the geographical distance matrix and categorical distance matrix, the server can inform all users their neighbor IDs. Although in DCLR, the risk of sending geolocation centroids and category distributions to the server is considered lower than directly transmitting raw user trajectories \cite{duriakova2019pdmfrec}, our framework fully accounts for the compatibility with privacy preservation mechanisms when communicating with the server. Specifically, we 
propose to inject noise into centroids and distributions to further reduce the privacy risk with the well-established differential privacy \cite{dwork2014algorithmic}. In DCLR, 
noise from Laplace distribution is injected into both centroids $O(u_n)$ and category distributions $CP(u_n)$ to achieve $\epsilon$-differential privacy:
\begin{equation}
  M(D) = f(D) + Lap(\frac{\Delta f}{\epsilon}) \label{dp:eq}
\end{equation}
\noindent{where} $D$ is the data to be submitted by a user, $f(\cdot)$ denotes an arbitrary randomized function, $\epsilon$ is the privacy budget, referring to the degree of privacy protection offered, 
and $\Delta f$ is the sensitivity of the function $f(\cdot)$, which is calculated as below:
\begin{equation}
  \Delta f = {max}_{D,D'}||f(D)-f(D')||_1
\end{equation}
\noindent{where $D$ and $D'$ are two adjacent datasets that differ in only one record.} 
Back to our scenario, for $O(u_n)$, the sensitivity can be easily obtained via the sum of longitude and latitude differences between two farthest centroids. For $CP(u_n)$, as mentioned above, category distribution is derived from user's visiting counts over all POI categories. In addition, 
the sensitivity of the counting query is 1 \cite{dwork2014algorithmic}. Thus, we can add noise satisfying $Lap(\frac{1}{\epsilon})$ distribution to the visiting counts of all categories before converting them into the category distribution. By adjusting the magnitude of privacy budget $\epsilon$, we can conveniently control the user privacy by trading off the informativeness of identified neighbors for each user.

\subsection{Collaborative Model Learning}
Given the geographical neighbors $N_{geo}(u_n)$ and semantic neighbors $N_{cat}(u_n)$, the new challenge 
is how to enhance the personalized model with both types of neighbors. As two different types of neighbors 
carry vastly different knowledge, simply aggregating all neighbors will cause considerable information loss, 
leading to a significant drop of recommendation quality. Hence, we propose a two-step method to solve the problem. 
Firstly, we generate two enhanced models by separately exchanging knowledge with geographical and semantic neighbors. 
Then, we further combine these two enhanced models into a fully personalized local recommender for each user. It is worth 
noting that such communication process is implemented locally while neighbors' model weights $\Theta$ are transferred 
device-wise from $N_{geo}(u_n)$ and $N_{cat}(u_n)$ to $u_n$.

In decentralized deep learning, the basic method to communicate with other homogeneous models is model aggregation \cite{zhang2021survey}. In our framework, both geographical and semantic neighbors have different weights as they have different distances to the current user. On this basis, we propose an affinity-based model aggregation strategy to learn knowledge from neighbors. To learn the enhanced model with geographical neighbors, we first calculate the similarity between a neighbor and the current user based on its distance to the current user:
\begin{equation}
  s(u_n, u_m) = \frac{1}{1+d_{geo}(u_n, u_m)}
\end{equation}
After that, we define the weight of each neighbor model $\Theta_m$ of $u_m\in N_{geo}(u_n)$ via:
\begin{equation}
  w(\Theta_n) = \frac{s(\Theta_m)}{\sum\limits_{k=1}^q s(\Theta_{k})}
\end{equation}
On this basis, the enhanced model for user $u_n$ with geographical neighbors is defined as below:
\begin{equation}
  \Theta_n^{geo} \leftarrow (1-\mu) \overline{\Theta}_n + \mu\!\sum\limits_{m=1}^q w(\Theta_{m})\,\Theta_{m} \label{abc}
\end{equation}
\noindent{where} $\overline{\Theta}_n$ is the original model for the current user and $\mu$ is a hyperparameter which controls the proportion of the current model and the aggregated model based on neighbors. 

The enhanced model $\Theta^{cat}_n$ with semantic/categorical neighbors can be done analogously to geographical neighbor aggregation, where $\mu$ is the 
same for both geographical neighbors and semantic neighbors. Given two enhanced models $\Theta^{geo}_n$ and $\Theta^{cat}_n$, 
another challenge is how to losslessly combine these two models to obtain the final 
model for each user. Both $\Theta^{geo}_n$ and $\Theta^{cat}_n$ are enhanced models, and they can be treated as two different views of the current model. 
Therefore, we adopt Mutual Information Maximization \cite{zhou2020s3} to let these two enhanced 
models gain knowledge from each other. Intuitively, the most important component of these two models is their embedded POI representations. Thus, we construct positive training samples by pairing the embeddings of the same POI from two models. For each positive pair, we 
generate negative samples by swapping one POI embedding with the embedding for a different POI randomly drawn from the corresponding model. Inspired by InfoNCE loss \cite{kong2019mutual}, 
we define the combination loss $L_{comb}$ for two models as below:
\begin{equation}
  L_{comb}= -\sum_{\forall p_i} \log\frac{\exp({f_{comb}(\textbf{e}^{geo}_{p_i}, \textbf{e}^{cat}_{p_i})})}{
  \sum\limits_{j=1,j\neq i}^{N_1} 
  \exp({f_{comb}(\textbf{e}^{geo}_{p_i}, \textbf{e}^{cat}_{p_j})})+\sum\limits_{j'=1,j'\neq i}^{N_2} \exp({f_{comb}(\textbf{e}^{geo}_{p_{j'}}, \textbf{e}^{cat}_{p_i})})}
\end{equation}
where $\textbf{e}^{geo}_{p_i}$ and $\textbf{e}^{cat}_{p_i}$ are embeddings of the same POI from two models, 
$\textbf{e}^{cat}_{p_j}$ and $\textbf{e}^{geo}_{p_{j'}}$ are respectively the negative embeddings from $\Theta^{cat}_n$ and $\Theta^{geo}_n$, 
$N_1=N_2$ are the numbers of negative samples from each of the models, 
and $f_{comb}(\cdot, \cdot)$ is the function of pairwise similarity implemented as the bilinear network below:
\begin{equation}
f_{comb}(\textbf{e}^{geo}_{p_i},\textbf{e}^{cat}_{p_i})=\sigma({\textbf{e}^{geo}_{p_i}}^T \textbf{W}_{comb} \textbf{e}^{cat}_{p_i})
\end{equation}
\noindent{where} $W_{comb}$ is a learnable parameter matrix. Once the finetuning of both enhanced models are finished by minimizing $L_{comb}$, we further average the two finetuned models to get the final personalized model: 
\begin{equation}
  \Theta_u \leftarrow \frac{(\Theta_u^{geo} + \Theta_u^{cat})}{2}
\end{equation}

To achieve above the process, users must collect model weights of their neighbors, which may lead to privacy breaches. To this end, users only share their perturbed weights. Similar to the strategy when sharing centroids and category distributions, we propose to inject noise satisfying Laplace distribution into model weights. During model weights exchange, the sensitivity of the model weights is $\frac{2\eta}{N_{pos}}$ where $\eta$ is the absolute difference between the largest and smallest weights of the model $\overline{\Theta}_u$, and $N_{pos}$ is the number of positive samples used to train the model \cite{wei2020federated}. Thus, for each user, we inject noise satisfying $Lap(\frac{2\eta}{N_{pos}\epsilon })$ to the model weights before transmitting them to neighboring user devices.

\renewcommand{\algorithmiccomment}[1]{\textbf{#1}}
\newcommand{\algorithmicnewcomment}[1]{\textit{#1}}

\begin{algorithm}[t]
  \caption{DCLR with similarity-based collaborative learning. Unless specified, processes are 
  implemented on user's side.}
  \label{alg:A}
\begin{algorithmic}[1]
  % \ENSURE A set of centers for the user
  \STATEx /*Server-side operation*/
    \STATE Initialize $\Theta;$\
    \REPEAT 
      \STATE Take a gradient step w.r.t. $L_{DP}$ to update $\Theta$;
      \STATE Take a gradient step w.r.t. $L_{CP}$ to update $\Theta$;
    \UNTIL{convergence; \hspace{4.5cm}$\triangleright$ Pretraining with self-supervision signals} 
    \STATE Sever receives perturbed $O(u_n)$ and $CP(u_n)$ from all $u_n\in U$;
  \STATE Calculate $\textbf{D}_{geo}$ with centroids $\{O(u_1),O(u_2),...,O(u_{|U|})\}$;
  \STATE Calculate $\textbf{D}_{cat}$ with category preferences $\{CP(u_1),CP(u_2),...,CP(u_{|U|})\}$;
  \STATE Send $\Theta$, $N_{geo}(u_n)$ and $N_{cat}(u_n)$ to each $u_n \in U$;
  \STATEx /*Server engagement ends*/
  \FOR[in parallel]{each $n=1,2,...,|U|$}
    	\STATE $\Theta_n \leftarrow \Theta$;
    \ENDFOR
  \REPEAT 
        \FOR[in parallel]{each $n=1,2,...,|U|$}
      \STATE Take a gradient step w.r.t. $L_{POI}$ to update $\Theta_n$, denoted as $\overline{\Theta}_n$; \,\,\,\,$\triangleright$ Local model optimization
    \ENDFOR  	
    \FOR[in parallel]{each $n=1,2,...,|U|$}
      \STATE $\Theta_n^{geo} \leftarrow (1-\mu) \overline{\Theta}_n + \mu\!\sum\limits_{m=1}^q w(\Theta_{m})\,\Theta_{m}$ with $N_{geo}(u_n)$;
      \STATE $\Theta_n^{cat} \leftarrow (1-\mu) \overline{\Theta}_n + \mu\!\sum\limits_{m'=1}^q w(\Theta_{m'})\,\Theta_{m'}$ with $N_{cat}(u_n)$;
      \STATE $\Theta_n^{geo}, \Theta_n^{cat} \leftarrow \textnormal{argmin}_{\{\Theta_n^{geo},\Theta_n^{cat}\}}
        L_{comb}$; \,\,\,\,\,\,\,\,\,\,\,\,\,\,\,\,\,\,\,\,\,\,\,$\triangleright$ Finetuning enhanced models via MIM
      \STATE $\Theta_n \leftarrow \frac{1}{2}(\Theta_n^{geo} + \Theta_n^{cat})$;
    \ENDFOR
  \UNTIL{convergence}
\end{algorithmic}
\end{algorithm}

\subsection{Learning Personalized Next POI Recommendation Model with DCLR}
The implementation of DCLR is described in Algorithm \ref{alg:A}. Starting from the server side, lines 1-5 
initialize the recommender and pretrain POI representations with two self-supervised learning signals. With 
perturbed centroid and category preferences received from all user devices (line 6), the server calculates the 
geographical distance matrix and categorical distance matrix via lines 7-8, then sends the neighbor information 
to all users (line 9). The server's engagement in the decentralized learning ends from here.
Back to the user's side, after obtaining the pretrained POI embeddings (lines 10-12), we record the updated 
user model parameters as $\overline{\Theta}_n$ in every training epoch (lines 14-16). Then, in 
lines 17-22, $\overline{\Theta}_n$ is enhanced by the models of geographically and semantically similar users, 
which will undergo a mutual information maximization step and be merged. The collaborative on-device learning 
process iterates until all local models converge or the maximum epoch number is reached.

\subsection{Complexity Analysis}
We now analyze the communication and computation complexities of Algorithm \ref{alg:A}. 
Recall that $q$ denotes the neighbor size for both geographical and semantic neighbors, $|P|$ denotes the number 
of POIs and $M$ denotes the maximum number of check-in activities. Besides, we use $E$ to denote the maximum number of epochs.

\textbf{Communication Complexity.} Before the training process, each user needs to upload their centroid $O(u_n)$
and category preferences $CP(u_n)$ to the server. After that, each user needs to receive the pretrained model $\Theta$, geographical 
neighbor ids $N_{geo}(u_n)$ and semantic neighbor ids $N_{cat}(u_n)$ from the server. During the training process, for each 
epoch, each user needs to collect $2q$ models from two types of neighbors. Simultaneously, each user needs to send 
out their model a certain number of times. The average times of sending out are $2q$ as the number of models collected 
should be the same as the number of models sent out. Thus, the total communication complexity of each 
user is $(4q + 1)*\Theta + O(u_n) + CP(u_n) + N_{geo}(u_n) + N_{cat}(u_n)$. The size of $\Theta$ is much larger than the 
sizes of $O(u_n)$, $CP(u_n)$, $N_{geo}(u_n)$ and $N_{cat}(u_n)$, and the most important component of $\Theta$ is POI embeddings. Thus, 
the total communication complexity for each user is linear with the POI set size $|P|$.

\textbf{Computation Complexity.} Since the parameter pretraining in DCLR is a one-off process 
performed on the server, we only analyze the computation complexity of the training process on-device. For each epoch, 
the computation cost mainly relies on two parts including local model optimization with personal trajectory and model enhancement 
via learning knowledge from neighbors. For local model optimization, the time complexity is $M$ since $M-2$ check-ins are used 
for training. For model enhancement, the first step is to obtain two enhanced models by aggregating models from two types of 
neighbors and the time complexity is $2q$. The second step is to combine the two enhanced models by MIM and the time complexity 
is $|P|$ since each POI embedding is a training sample. Thus, the total computation complexity of each user is $E(M+2q+|P|)$. 
Since $M$ and $|P|$ are relatively larger than $q$, the total computation complexity of each user mainly depends on 
the number of training samples on-device and the POI set size.

The above complexity analysis shows that DCLR is efficient and can be scaled to very large datasets.

\section{EXPERIMENTS}
In this section, we evaluate the recommendation performance of DCLR in POI 
recommendation w.r.t. state-of-the-art baselines. All experiments are processed on the computer equipped with 
Intel (R) Core i7-10700K CPU, 3.80GHz and one NVIDIA's GeForce RTX 2070, 8GB Graphics Card.

\subsection{Datasets and Evaluation Protocols}
We evaluate our proposed DCLR on two publicly available real-world Location-Based Social Network 
datasets: Weeplace \cite{liu2013personalized} and Yelp \cite{DBLP:journals/corr/FanK14}. 
Both datasets contain user check-ins in the cities of New York, Los Angeles and Chicago.
Inspired by previous works \cite{chang2018content,Li2018NextPR}, we remove users with less than 10 
check-ins for both datasets, as well as POIs that have less than 10 visits.
Table \ref{table:A} summarizes the statistics of the two datasets.

% Table 1: Dataset statistics.
\begin{table}[htbp]
  \centering
  \caption{Dataset statistics.}
  \label{table:A}
   \vspace{0.1mm}
    \begin{tabular}{lrr}
          \hline
          & \multicolumn{1}{r}{Weeplace} & \multicolumn{1}{r}{Yelp} \\
    \hline
    \#users & 4504     & 6070 \\
    \hline
    \#POIs & 35675     & 46527 \\
    \hline
    \#categories & 450     & 428 \\
    \hline
    \#check-ins & 886408     & 951832 \\
    \hline
    \#check-ins per user & 196.80   & 156.81 \\
    \hline
    \end{tabular}%
  % \label{tab:addlabel}%
\end{table}%

% typo checked
For evaluation, we adopt the leave-one-out protocol which is widely used in previous work 
\cite{wang2018neural, wang2019enhancing}. That is, for each check-in sequence, we use the 
last check-in POI for test, the second last POI for validation, and the remaining for training. 
The maximum sequence length is set to 200. The most recent 200 check-ins will be used in the 
evaluation if the sequence length is larger than 200.
For each ground truth, we pair 
it with 200 unvisited POIs  as the candidates for ranking. Those 200 POIs are the closest ones to 
the user's last check-in location. Note that, different from ranking e-commerce products \cite{krichene2020sampled,8509453}, our evaluation is based on the location-sensitive nature 
of POI recommendation tasks, as a user rarely travels long distances between two consecutive 
check-ins. 
Then, the local recommendation model generates a ranked list for those 201 POIs based on their scores, and 
the ground truth is expected to be ranked at the top. Then, we adopt Hit 
Ratio at Rank $k$ (HR@$k$) and Normalized Discounted Cumulative Gain at Rank $k$ (NDCG@$k$) \cite{weimer2007cofi} to measure 
the recommendation on the top-$k$ POIs in the ranking list, where HR@$k$ calculates the times 
that the ground truth is among the top-$k$ list, while NDCG@$k$ further considers the ground truth's position 
in the top-$k$ list. 

% typo checked
\subsection{Baseline Methods}
We compare DCLR with the following baseline methods, including both centralized POI recommenders 
and state-of-the-art on-device ones:
\begin{itemize}
  \item \textbf{POP} is a simple method that ranks the POIs based on their popularity.

  \item \textbf{DIS} is a basic method for POI recommendation that ranks the POIs based on 
  their distances to the user's histrocial check-in POIs.

  \item \textbf{MF \cite{10.1145/2623330.2623638}} is a classic collaborative filtering algorithm for 
  centralized POI recommender systems.

  \item	\textbf{LSTM \cite{hochreiter1997long}} is a recurrent neural network that has shown its 
  ability to capture both short-term and long-term dependencies in sequential data. Thus, LSTM is 
  naturally applied to solve next POI recommendation problem.
  
  \item	\textbf{STAN \cite{LuoYingtao2021SSan}} is a state-of-the-art model that explicitly exploits 
  relative spatiotemporal information of both consecutive and non-consecutive check-in activities.
  It proposes a bi-attention architecture for personalized item frequency where the first layer 
  aggregates spatiotemporal information and the second layer matches the target to all check-ins.
  
  \item	\textbf{DMF \cite{chen2018privacy}} is a decentralized MF framework. To 
  overcome the data sparsity issues, on-device models learn knowledge from neighbors who 
  are physically close to the current user decided by the random walk theory. 
  
  \item	\textbf{LLRec \cite{WangQinyong2020NPRo}} is an on-device framework that adopts the 
  teacher-student training strategy. It trains a teacher RNN-based model with public data on the 
  cloud and sends the compressed model to the end devices for local model training. In this way, 
  end devices keep private check-in data without exposure to the cloud and the simplified model 
  significantly reduces the computation burden for end devices.

  \item	\textbf{PREFER \cite{GuoYeting2021PPRw}} is a federated learning framework for next 
  POI recommendation. Firstly, users train personalized models locally with their individual 
  check-in data in parallel. After that, the edge services collect and aggregate multi-dimensional 
  user-independent parameters to build a federated POI recommendation model. Then, the federated 
  model is sent back to users.
  
\end{itemize}  
  
For DCLR's hyperparameters, we set the latent dimension $d=32$, the neighbor size $q=30$ 
for both geographical and semantic neighbors, the trade-off between the current model 
and aggregated model of neighbors $\mu=0.3$ for both geographical and semantic neighbors, 
and the privacy budget $\epsilon=0.1$. The impacts of the above hyperparameters will be 
further discussed in Section 4.5. During training, we adopt a learning rate of $0.002$, dropout of $0.2$ on all deep layers, batch size of $16$, and the maximum training epoch is set to $50$.  
In addition, the numbers of negative samples for POI prediction loss ($N_{neg}$), category 
prediction LOSS ($N_{CP}$) and combination loss ($N_1=N_2$) are set to 5. For all baselines, we adopt the same general hyperparameters including latent dimension, learning rate, dropout, batch size and the maximum training epoch.

\subsection{Recommendation Performance}
% Recommendation performance comparison with baselines

% Table generated by Excel2LaTeX from sheet 'Sheet3'
\begin{table*}[t]
  \centering
  \caption{Recommendation performance comparison with baselines.}
  \label{table:B}
  \resizebox{\linewidth}{!}{
    \begin{tabular}{|l|cccc|cccc|}
      \hline
          & \multicolumn{4}{c|}{Weeplace}  & \multicolumn{4}{c|}{Yelp} \\
          \cline{2-9}
          & \multicolumn{1}{c}{HR@5} & \multicolumn{1}{c}{NDCG@5} & \multicolumn{1}{c}{HR@10} & \multicolumn{1}{c|}{NDCG@10} & \multicolumn{1}{c}{HR@5} & \multicolumn{1}{c}{NDCG@5} & \multicolumn{1}{c}{HR@10} & \multicolumn{1}{c|}{NDCG@10} \\
          \hline
          POP   & 0.0461 & 0.0328 & 0.0724 & 0.0545 & 0.0242 & 0.0168 & 0.0489 & 0.0336 \\
          DIS   & 0.0691 & 0.0453 & 0.1015 & 0.0739 & 0.0491 & 0.0352 & 0.0701 & 0.0514 \\
          MF    & 0.1493 & 0.1035 & 0.1966 & 0.1387 & 0.1312 & 0.0897 & 0.1534 & 0.0996 \\
          LSTM  & 0.3368 & 0.2256 & 0.4430 & 0.2926 & 0.2763 & 0.1893 & 0.3765 & 0.2399 \\
          STAN  & 0.4398 & 0.3017 & 0.5512 & 0.3818 & 0.4155 & 0.2889 & 0.5201 & 0.3521 \\
          DMF   & 0.1741 & 0.1577 & 0.2374 & 0.1984 & 0.1769 & 0.1125 & 0.1991 & 0.1252 \\
          LLRec & 0.4117 & 0.2733 & 0.5022 & 0.3421 & 0.4124 & 0.2726 & 0.4935 & 0.3461 \\
          PREFFER & 0.4285 & 0.2989 & 0.5297 & 0.3761 & 0.4263 & 0.2880 & 0.5180 & 0.3595 \\
          DCLR & \textbf{0.4452} & \textbf{0.3083} & \textbf{0.5569} & \textbf{0.3861} & \textbf{0.4388} & \textbf{0.3031} & \textbf{0.5466} & \textbf{0.3794} \\
          \hline
        \end{tabular}%
}
  \label{tab:addlabel}%
\end{table*}%

\begin{table}[t!]
  \centering
  \caption{Performance comparison between DCLR and STAN with multiple models (HR@10 is demonstrated).}
    \begin{tabular}{|l|c|c|c|}
          \hline
          & \multicolumn{1}{c|}{New York} & \multicolumn{1}{c|}{Los Angeles} & \multicolumn{1}{c|}{Chicago} \\
    \hline      
    STAN - single model & 0.5218 & 0.5162 & 0.5208 \\
    \hline
    STAN - multiple models & 0.5379 & 0.5199 & 0.5303 \\
    \hline
    DCLR  & 0.5494 & 0.5394 & 0.5483 \\
    \hline
    \end{tabular}%
  \label{table:D}%
\end{table}%

% typo checked
Table \ref{table:B} summarizes the experimental results on recommendation effectiveness. From such 
results, we have the following observations. 
Among the centralized POI recommendation, POP and DIS have poor performance on both datasets since they cannot capture users' preferences. 
After that, the fact that LSTM outperforms MF shows LSTM's ability to capture both short-term and long-term dependencies on 
next POI recommendation. In addition, thanks to the effective use of spatiotemporal information of both 
consecutive and non-consecutive check-in activities, 
STAN shows higher accuracy than LSTM. 
Compare to the advanced centralized baseline STAN, our proposed model also has competitive performance. 
Intuitively, on Weeplace, DCLR achieves light improvements over STAN (1\% on average for 
all metrics). On Yelp, DCLR outperforms STAN by 5\% to 8\% on all metrics. For this result, a possible reason is 
that, STAN is trained with check-ins across three different cities where the information learned from other cities 
might be noisy, leading to inferior performance of STAN. With this observation, we further show the effectiveness 
of DCLR as users under DCLR only exchange knowledge with similar users without overfitting the noise from dissimilar 
users. On this basis, for Yelp, we train and test three STAN models separately with check-ins from each of the three 
cities in the dataset. Then, we compare their test results with DCLR's performance over three cities separately. It 
is worth noting that DCLR is trained with the whole Yelp dataset and personalized models can still exchange knowledge 
with semantic neighbors from the other two cities. In addition, we also evaluate the performance STAN when a unified 
model is trained over three cities. The results are shown in Table 3, where we can observe noticeable performance 
improvements for all cities if three STAN models are trained separately rather than only one STAN model is used, 
which proves the above assumption. Even so, DCLR outperforms city-specific STAN models by 2\%, 4\% and 3\% in New York, 
Los Angeles and Chicago respectively. This is because DCLR can achieve better personalization, and learn more 
expressive models with the collaborative learning architecture.
Considering that DCLR significantly reduces the dependency on the cloud server 
and the risk of privacy breaches, the above results have demonstrated the effectiveness of our proposed 
method.

\begin{table*}[t!]
  \centering
  \caption{Ablation study. We mark DCLR's performance with boldface numbers.}
  \label{table:C}
  \resizebox{\linewidth}{!}{
    \begin{tabular}{|l|cccc|cccc|}
    \hline
          & \multicolumn{4}{c|}{Weeplace}  & \multicolumn{4}{c|}{Yelp} \\
    \cline{2-9}
          & \multicolumn{1}{c}{HR@5} & \multicolumn{1}{c}{NDCG@5} & \multicolumn{1}{c}{HR@10} & \multicolumn{1}{c|}{NDCG@10} & \multicolumn{1}{c}{HR@5} & \multicolumn{1}{c}{NDCG@5} & \multicolumn{1}{c}{HR@10} & \multicolumn{1}{c|}{NDCG@10}\\
    \hline
    DCLR & \textbf{0.4452} & \textbf{0.3083} & \textbf{0.5569} & \textbf{0.3861} & \textbf{0.4388} & \textbf{0.3031} & \textbf{0.5466} & \textbf{0.3794} \\
    -CP    & 0.4247 & 0.2898 & 0.5253 & 0.3584 & 0.4179 & 0.2821 & 0.5139 & 0.3593 \\
    -DP    & 0.4119 & 0.2801 & 0.5027 & 0.3498 & 0.4035 & 0.2700 & 0.4951 & 0.3433 \\
    -AN    & 0.2819 & 0.1948 & 0.3183 & 0.2450 & 0.2632 & 0.1858 & 0.3491 & 0.2252 \\
    -GN    & 0.3432 & 0.2430 & 0.3959 & 0.3033 & 0.3289 & 0.2242 & 0.4282 & 0.2746 \\
    -SN    & 0.3743 & 0.2558 & 0.4135 & 0.3199 & 0.3401 & 0.2426 & 0.4534 & 0.2979 \\
    -MIM   & 0.4164 & 0.2829 & 0.5251 & 0.3741 & 0.3896 & 0.2724 & 0.5123 & 0.3597 \\
    -PP    & 0.4618 & 0.3214 & 0.5746 & 0.4059 & 0.4400 & 0.3074 & 0.5648 & 0.3878 \\
    \hline
    \end{tabular}%
  }
  \label{tab:addlabel}%
\end{table*}%

% \begin{table}[t!]
%   \centering
%   \caption{Performance comparison between DCLR and STAN with multiple models (HR@10 is demonstrated).}
%     \begin{tabular}{|l|c|c|c|}
%           \hline
%           & \multicolumn{1}{c|}{New York} & \multicolumn{1}{c|}{Los Angeles} & \multicolumn{1}{c|}{Chicago} \\
%     \hline      
%     STAN-single model & 0.3532 & 0.3495 & 0.3526 \\
%     \hline
%     STAN-multi models & 0.3733 & 0.3539 & 0.3680 \\
%     \hline
%     DCLR  & 0.3813 & 0.3744 & 0.3805 \\
%     \hline
%     \end{tabular}%
%   \label{table:D}%
% \end{table}%

% typo checked
In addition, our method consistently and significantly outperforms all three on-device POI 
recommendation models in terms of every metric on both datasets. In detail, our work shows 
the advantage against DMF by an obvious margin, where the improvement achieves 147\% on average 
for all metrics. This is because our work exploits multiple components to handle the data sparsity 
issues. Firstly, We propose two self-supervised learning objectives to inject geographical and 
categorical information into POI representations. Then, we propose an effective way to learn 
knowledge from two types of neighbors while DMF only learns knowledge from 
geographical neighbors. 
Furthermore, our proposed model outperforms LLRec by 10\% on average for all metrics. To completely 
protect users' privacy, LLRec does not allow any form of personal information to leave users' devices. 
Instead, our proposed method is able to bring large improvement on the recommendation quality with 
negligible privacy risks given the information shared and the differential privacy guarantee.
Meanwhile, both the teacher model and the compressed model are trained on the cloud, showing high 
demand of on-cloud computing and storage resources. Thus, DCLR has shown its advantages over LLRec.
Compared to PREFFER, our work has a noticeable advantage on recommendation accuracy. 
In addition, PREFFER needs to collect and aggregate users' personalized models, exposing user 
privacy to potential breaches. Therefore, DCLR is more capable than PREFFER since it can provide 
both stronger privacy protection and more accurate recommendations with less reliance on the cloud 
server. 

% typo checked
\subsection{Ablation Study}
To better understand the performance gain from various components of our methods, we conduct an 
ablation study on different degraded versions of DCLR. Table \ref{table:C} summarizes the results. 
We denote the full model as DCLR and drop different components as variants. 
In what follows, we introduce all variants and discuss the effectiveness of responding model 
components.

\textbf{Remove Category Prediction Loss (CP)}. In DCLR, we exploit category prediction loss and 
distance prediction loss to infuse categorical and geographical information into POI representations. 
To testify the usefulness of category prediction loss, we remove category prediction loss and only 
distance prediction loss is used to enhance POI representations. As a result, we observe performance 
drops (6\% on average for all metrics) for both datasets. This is because users' interactions at the 
category level are relatively denser compared to the POI level, helping the model capture the preferences 
on POIs.

\textbf{Remove Distance Prediction Loss (DP)}. To evaluate the effectiveness of distance prediction 
loss, we remove distance prediction loss and only category prediction loss is used to enhance POI 
representations. Consequently, there is a noticeable decrease in recommendation performance 
(9\% on average for all metrics) for both datasets. Hence, the inclusion of distance prediction loss 
is necessary since it can help to capture transformation rules among sequential check-in activities. 
In addition, representative POI embeddings with categorical and geographical information are 
necessary for personal models to exchange knowledge from neighbors with highly relevant patterns. 
As such, the two self-supervised losses have shown their strong effectiveness to improve the 
recommendation quality.

\textbf{Remove All Neighbors (AN)}. The crucial component of DCLR is allowing users to exchange 
knowledge with two types of neighbors, which is supposed to address the data sparsity issues caused 
by the fact that all personalized models are trained locally with individual check-in data. To 
verify its efficacy, we remove the neighbor-based aggregation for both types of neighbors and each 
user learns a single local recommendation model with his/her own data. Consequently, 
the collaborative learning component and privacy protection are also removed. 
The resulted variant model has experienced serious performance drops (39\% on average for all metrics) for both datasets.
Among all components of DCLR, the performance drops of removing all neighbors are the most 
significant. Such results have shown that our strategy of exchanging knowledge with both types of 
neighbors can effectively handle the data sparsity issues in the decentralized environment. 
Meanwhile, this strategy also avoids learning biased information from irrelevant users. 

\textbf{Remove Geographical Neighbors (GN)}. Although the combination of both types of neighbors has 
proven its effectiveness, we are still interested in the performance gain from the single neighbor
type. To testify the effectiveness of geographical neighbors, we only allow local models to 
exchange knowledge with semantic neighbors, and thus, we dont need to combine knowledge of two types 
of neighbors via mutual information maximization. As a consequence, significant performance drops 
(24\% on average for all metrics) for both datasets are observed, showing the effectiveness of 
exchanging knowledge with geographical neighbors.

\textbf{Remove Semantic Neighbors (SN)}. To verify the usefulness of semantic neighbors, 
we only allow local models to learn knowledge from geographical neighbors. Similarly, mutual information 
maximization for aggregating knowledge from two types of neighbor is removed. Accordingly, we observe obvious 
performance drops (20\% on average for all metrics) from both datasets. As such, the knowledge learned from 
semantic neighbors is indispensable for addressing the data sparsity issues. In addition, considering the 
performance of removing geographical neighbors, it can be proven that geographical neighbors can provide more 
valuable information compared with semantic neighbors. Besides, as mentioned above, distance prediction is a 
more influential self-supervised task compared with category prediction. Thus, we can conclude that, in POI 
recommendation, although semantic factors significantly affect the recommendation performance, geographical 
factors do play more important roles. On this basis, in the future research and deployment of decentralized 
POI recommendation systems, resources can be slightly skewed to geographical factors (e.g., assigning more 
geographical neighbors).

\textbf{Remove Mutual Information Maximization (MIM)}.
Given two enhanced models learned separately from two types of neighbors with the attention mechanism, 
we design a mutual information maximization task to further combine these enhanced models into 
a fully personalized local model. To verify the effectiveness of this task, we simply use the 
average method rather than MIM to combine the two enhanced models. As a consequence, we observe 
a noticeable decrease of recommendation performance (7\% on average for all metrics) for both 
datasets. Thus, the inclusion of MIM is important to the information fusion of two different 
types of neighbors.

However, the MIM process is fully performed on-device and high 
computational overhead might crash the devices. To verify whether the computational overhead 
added is reasonable to low-power devices, we record the training time where MIM is not included and 
MIM is included. Please note the time is the average time per user per epoch. 
The results are shown in Figure \ref{fig:something2}. We can observe significant increases of training time if MIM is included. 
This is because all POI embeddings are updated during the MIM process. However, the MIM component is able 
to significantly improve each local model's recommendation accuracy with only around 6 seconds' time added 
to the aggregation step on top of its plain counterpart without MIM. Furthermore, as MIM is performed 
during the model aggregation, this does not affect the inference time of the final model.

\begin{figure*} 
	\centering 
	\includegraphics[width=4in]{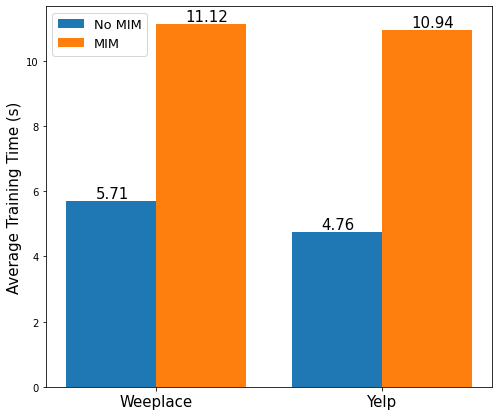}
	\caption{Computation Overheads of MIM Component.}
	\label{fig:something2} %% label for entire figure 
\end{figure*}

\textbf{Remove Privacy Protection (PP)}. In DCLR, we exploit differential privacy mechanism 
to avoid privacy breaches during the processes of neighbor identification and communication. 
To measure the influence of differential privacy mechanism on the performance of DCLR, we remove 
the privacy protection. After that, the performance is slightly improved by 3\% on average 
for all metrics and both datasets. This reflects that our model can provide strong privacy 
protection with only a little sacrifice in recommendation quality.

\begin{figure*} 
	\centering 
	\includegraphics[width=5.4in]{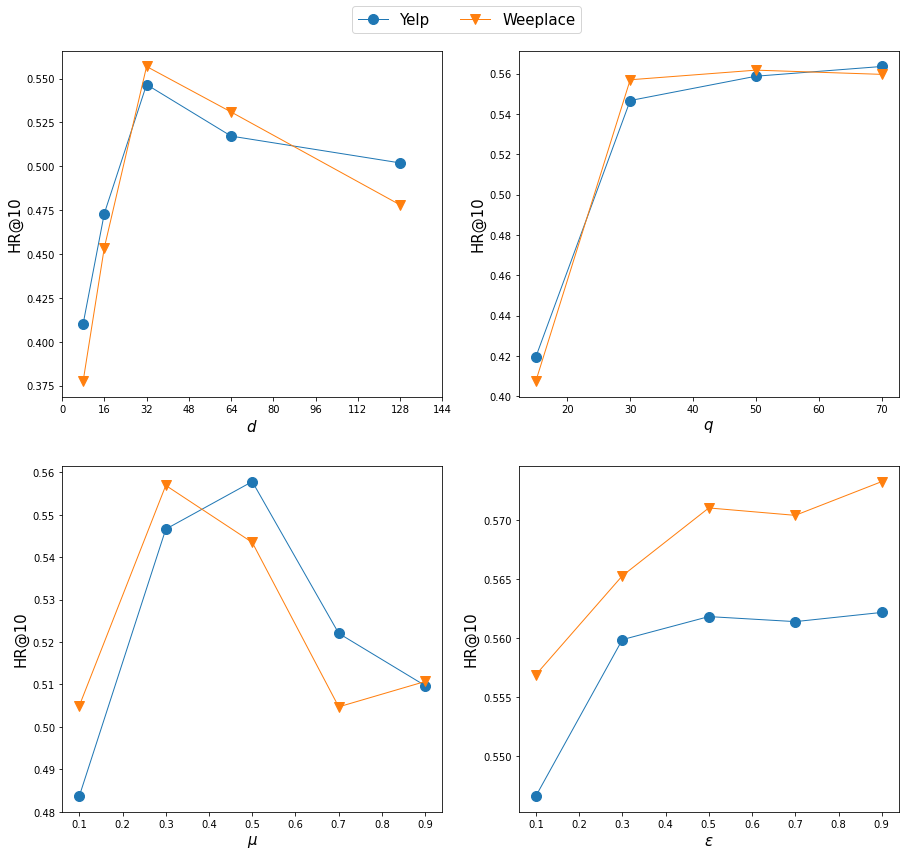}
	\caption{Recommendation performance with different hyperparameters.}
	\label{fig:something} %% label for entire figure 
\end{figure*}
\begin{figure*} 
	\centering 
	\includegraphics[width=5.4in]{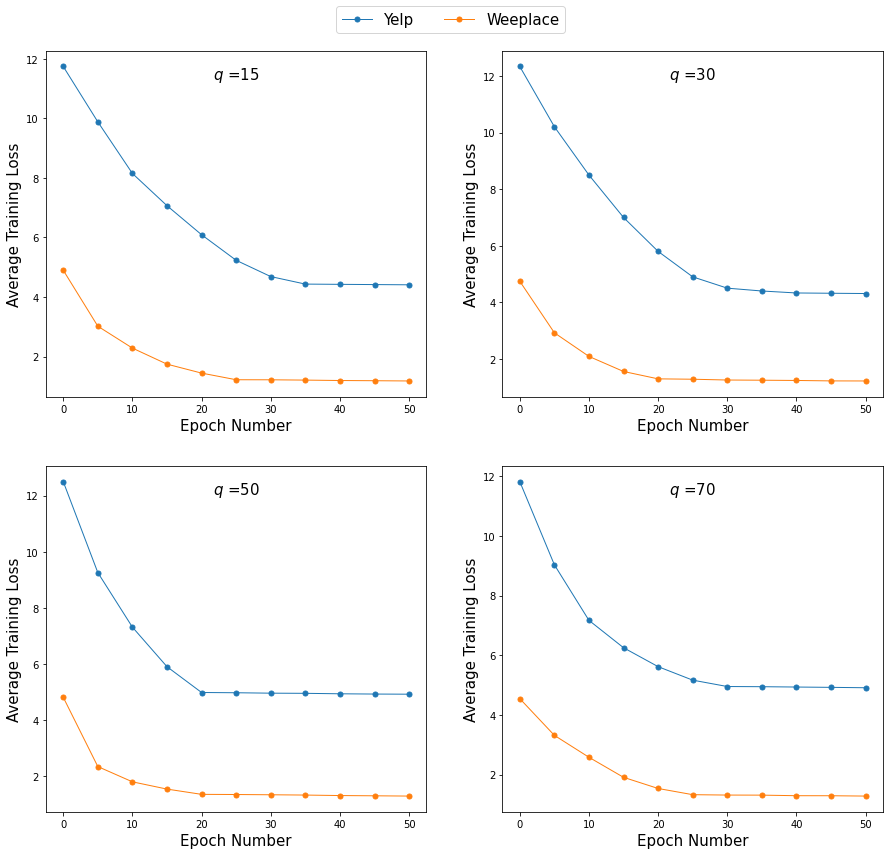}
	\caption{Effect of neighbor sizes on convergence speed.}
	\label{fig:something1} %% label for entire figure 
\end{figure*}

% typo checked
\subsection{Hyperparameter Sensitivity}
In this section, we investigate the performance fluctuations of our proposed DCLR with varying 
hyperparameters including the latent dimension $d$, the neighbor size $q$ used for both geographical and 
semantic neighbors, the trade-off between the current model and the aggregated model $\mu$ in Eq.(\ref{abc}), 
and the privacy budget $\epsilon$. The results are shown in Figure \ref{fig:something}.

\textbf{Impact of $d$.} Firstly, we study the impact of the latent dimension $d\in\{8,16,32,64,128\}$. 
As $d$ increases from 8 to 32, 
significant improvements of DCLR's performance are observed for both datasets. However, when $d$ is larger 
than 32, the performance starts to decrease. A possible reason is that, in DCLR, personalized models are 
trained with limited data and larger dimension might cause overfitting problems.

\textbf{Impact of $q$.} Then, we experiment on a series of neighbor sizes $q\in\{15,30,50,70\}$. 
Generally, DCLR benefits from larger $q$ on both datasets. However, the performance growth tends 
to slow down when $q$ is large enough (50 and 70). In addition, the drop of accuracy (Weeplace) can 
be also observed. The reason is that neighbors who have larger geographical distances and smaller 
categorical preference similarities contain biased information. Moreover, the increase of $q$ also 
leads to huge computing resources. Thus, the neighbor size should be controlled within a reasonable 
range.

\textbf{Impact of $\mu$.} $\mu$ is a hyperparameter which controls the proportion of the current 
model and the aggregated model. The value of $\mu$ is examined in $\mu\in\{0.1, 0.3, 0.5, 0.7, 0.9\}$. 
On Weeplace, DCLR achieves the best performance when $\mu=0.3$. However, on Yelp, DCLR achieves the 
best performance when $\mu=0.5$. This is because the users of the Yelp dataset have fewer check-in 
activities, leading to more serious data sparsity problems. Hence, more information from the 
neighbors will greatly help improve the performance of the final local models.

\textbf{Impact of $\epsilon$.} $\epsilon$ is the privacy budget which controls the amount of noise 
added to centroids, category distributions and model weights. The higher $\epsilon$ is, the less 
noise is added. We evaluate the impact of different privacy budgets 
$\epsilon\in\{0.1,0.3,0.5,0.7,0.9\}$. As the privacy budget increases from 0.1 to 0.9, there is 
generally an upward trend in DCLR's performance, but the improvement tends to stop when $\epsilon$ 
is larger than 0.5. The reason is that , when $\epsilon$ is relative large, the limited amount of 
noise will have minor side effects on the recommendation accuracy. But meanwhile, a lower noise 
injection will be less effective in protecting users' privacy. In addition, the noise added to 
model weights can be regarded as the regularization term which can effectively prevent overfitting. 
Hence, by default, $\epsilon$ is set to 0.1 which can provide strong privacy protection with 
acceptable sacrifice of accuracy.

% \textbf{Impact of $\epsilon$.} I love u.

% \begin{figure*} 
% 	\centering 
% 	\includegraphics[width=6in]{output3.png}
% 	\caption{Recommendation performance with different hyperparameters}
% 	\label{fig:something} %% label for entire figure 
% \end{figure*}
% \begin{figure*} 
% 	\centering 
% 	\includegraphics[width=6in]{output4.png}
% 	\caption{Effect of Different Neighbors on Convergence Speed}
% 	\label{fig:something1} %% label for entire figure 
% \end{figure*}

% typo checked
\subsection{Effect of Neighbor Sizes on Convergence Speed}
In decentralized machine learning, convergence speed is a crucial performance metric, as it 
determines the efficiency of the model deployment, as well as the ability to ensure timeliness 
when processing new data. In this work, we further evaluate the convergence speed on various 
neighbor sizes $q\in\{15,30,50,70\}$ by displaying the average training loss on iteration number. 
The results are shown in Figure \ref{fig:something1}. The first observation is that, among all 
neighbor sizes, the convergence speed of DCLR on Weeplace is faster than that of Yelp, as Weeplace 
data is relatively denser compared to Yelp. Secondly, when the neighbor size is increased 
from 15 to 30, the convergence speeds have experienced a noticeable rise. This is because, with 
more neighbors, personalized models can obtain richer information to capture users' preferences 
more quickly. However, as the neighbor size continues to rise, the quality of neighbors become 
worse since those neighbors have larger geographical distances and smaller categorical preference similarities. 
Thus, knowledge learned from such neighbors is biased which reduces the efficiency of learning 
users' preferences. As a result, with the increase of neighbor size, the convergence speed 
increases at first. Then, the convergence will gradually slow down as the neighbor size continues 
to increase further.

% typo checked
\section{RELATED WORK}
In this section, we review recent literature on related areas including POI 
recommendation and decentralized learning.
% typo checked
\subsection{Next POI recommendation}
Similar to traditional item recommenders, early POI recommender systems exploit the 
collaborative filtering technique to capture the correlations among users, POIs and 
contextual features \cite{10.1145/2623330.2623638,10.5555/2540128.2540504}.
Intuitively, GeoMF \cite{10.1145/2623330.2623638} captures the spatial clustering phenomenon in terms 
of two-dimensional kernel density estimation by augmenting users' and POIs' latent factors 
in the factorization model with activity area vectors of users and influence area vectors 
of POIs, respectively. FPMC-LR \cite{10.5555/2540128.2540504} extends the learning of latent features 
to personalized Markov chains and the localized regions. As a result, it can exploit the personalized 
Markov chain in the check-in sequences and takes into account users' movement constraints.

Recent POI recommendation models are mainly based on recurrent neural networks (RNNs) and their variants,
which achieve the state-of-the-art performance. 
To improve model performance, ST-RNN \cite{10.5555/3015812.3015841} models local temporal and spatial 
contexts in each layer with time-specific transition matrices for different time intervals and 
distance-specific transition matrics for different geographical distances.
SERM \cite{yao2017serm} combines spatiotemporal information and semantic contexts (e.g., keywords and 
categorical tags) to reflect user preferences and it effectively captures semantics-aware spatiotemporal 
transition regularities to improve POI prediction accuracies.
However, the above RNN models ignore long-term dependencies within the check-in 
trajectories. To solve this problem, DeepMove \cite{10.1145/3178876.3186058} firstly uses a recurrent 
layer to learn short-term sequential regularity from highly correlated trajectories, and then combines 
it with long-term periodicity learned by an attention layer. STGN \cite{9133505} adds time and distance 
gates to the standard LSTM  for better capturing short-term and long-term spatiotemporal preference.
To further improve the performance of LSTM, ARNN \cite{guo2020attentional} models both the sequential 
regularity and transition regularities of similar POIs to build the knowledge graph. 

Inspired by sequential item recommendation \cite{kang2018self}, GeoSAN \cite{lian2020geography} utilizes 
self-attention to capture user preference with point-to-point interactions among the trajectory. 
To make use of correlations of non-adjacent POIs and non-consecutive check-ins, STAN \cite{luo2021stan} 
uses a bi-layer attention architecture where the first layer aggregates spatiotemporal correlations 
within the trajectory and the second layer recalls the target with consideration of personalized item 
frequency (PIF).

Nevertheless, all the above methods are trained and deployed on cloud services, causing significant 
problems including huge cost of resources, privacy issues and high demand of powerful networks. Instead, 
our proposed DCLR deploys personalized models on mobile devices which can stably provide accurate and 
secure POI recommendations.

% typo checked
\subsection{On-device frameworks for next POI reccommendation}
On-device frameworks appear to overcome most shortcomings of centralized learning, and it 
has been applied to many areas such as multiarmed bandit \cite{kalathil2014decentralized}, 
network distance prediction \cite{liao2010network}, hash function learning \cite{leng2015hashing} 
and health analysis \cite{ye2021personalized}. 
There are also multiple works to deploy on-device POI recommendation models. Firstly, 
Chen et al. \cite{chen2018privacy} proposed a decentralized matrix factorization framework (DMF) for 
next POI recommendation, where personalized models are trained and stored locally. To improve the 
recommendation precision, DMF exchanges knowledge with neighbors who are physically close to the 
current user.
Then, Wang et al. \cite{WangQinyong2020NPRo} deployed compressed models on mobile devices for secure 
and stable POI recommendations. To maintain the robustness of the whole on-device framework, 
local compressed models inherit the knowledge from the teacher model which is trained with public data. 
Nevertheless, all users share the same model trained with the public data, ignoring the dynamics 
and diversity of users' spatial activities and interests. Instead, our proposed work allows users to obtain 
high-quality, fully personalized local recommenders by collaboratively learning knowledge with other similar 
users.
In addition, Guo et al. \cite{GuoYeting2021PPRw} proposed a federated learning framework for 
next POI recommendation. That is, personalized models are trained and stored locally. Then, the edge 
servers collect and aggregate all personalized models. After that, the aggregated model is sent 
back to all users. Consequently, the federated framework is still resource-intensive for storage and 
computation. Besides, all users in the federated framework also share the same global model. However, in our proposed 
work, the central server is only responsible for providing pretrained parameters and grouping similar users in a 
secure way, and local models are full personalized via an efficient collaborative learning strategy.

% typo checked
The most similar work to ours is the decentralized matrix factorization for next POI recommendation 
\cite{chen2018privacy}. We summarise two major differences. 
(1) To alleviate the sparsity issues, We propose two self-supervised learning objectives to enhance POI 
representations with geographical coordinates and categorical tags. Specifically, the first task is to 
create correlations between POIs and their associated categorical tags by mutual information maximization, 
and the second is to learn and predict geographical distances between POIs.
(2) Their personalized models are enhanced by knowledge from users who are physically close to the 
current user. However, such limited knowledge cannot effectively address the data sparsity problem. 
On the contrary, we propose two types of neighbors based on geographical distances and categorical 
preference similarities. Then, we design a mutual information maximization task to jointly learn and 
combine knowledge from both types of neighbors in an effective way.

% typo checked
\section{CONCLUSION}
Advanced performance of current next POI recommenders is highly dependent on the collection, storage, 
and training that involve massive check-in data, leading to problems including huge cost of computing resources, 
privacy issues and high demand of network connectivity. In our paper, the proposed solution DCLR is a 
decentralized paradigm, that locally trains personalized recommenders for all users. To address the data 
sparsity issues, we firstly design two self-supervised signals to enhance POI representations with coordinate 
information and categorical tags. Then, for decentralized collaborative learning, we define two metrics to 
identify neighbors concerning geographical distances and categorical preference similarities. Finally, 
we exploit attention mechanism and mutual information maximization technology to jointly learn and combine 
knowledge from both types of neighbors. 
We evaluate the proposed DCLR with two real-world datasets. The experimental results have demonstrated 
our model's superiority in next POI recommendation. This is because, compared to state-of-the-art POI 
recommenders, it can provide more accurate personalized POI recommendations and stronger privacy protection  
with less reliance on the cloud server.
Through the ablation study and sensitivity analysis, we also show the significant effect of the two 
self-supervised learning tasks, the strategies for neighbor identification and communication, and the 
privacy protection mechanism at addressing data sparsity issues and providing privacy protection. 
Future work can include the fusion of social relationships for neighbor identification, and dynamic addition 
and removal of users with respect to timestamps.

\section*{ACKNOWLEDGEMENT}
This work is supported by Australian Research Council Future Fellowship (Grant No. FT210100624), Discovery Project (Grant No. DP190101985) and Discovery Early Career Research Award (Grant No. DE200101465). 

%\begin{acks}
%To Robert, for the bagels and explaining CMYK and color spaces.
%\end{acks}

\bibliographystyle{ACM-Reference-Format}
\bibliography{sample-acmsmall}


\begin{thebibliography}{59}

%%% the Web address.

\ifx \showCODEN    \undefined \def \showCODEN     #1{\unskip}     \fi
\ifx \showDOI      \undefined \def \showDOI       #1{#1}\fi
\ifx \showISBNx    \undefined \def \showISBNx     #1{\unskip}     \fi
\ifx \showISBNxiii \undefined \def \showISBNxiii  #1{\unskip}     \fi
\ifx \showISSN     \undefined \def \showISSN      #1{\unskip}     \fi
\ifx \showLCCN     \undefined \def \showLCCN      #1{\unskip}     \fi
\ifx \shownote     \undefined \def \shownote      #1{#1}          \fi
\ifx \showarticletitle \undefined \def \showarticletitle #1{#1}   \fi
\ifx \showURL      \undefined \def \showURL       {\relax}        \fi
% invisible to TeX
\providecommand\bibfield[2]{#2}
\providecommand\bibinfo[2]{#2}
\providecommand\natexlab[1]{#1}
\providecommand\showeprint[2][]{arXiv:#2}

\bibitem[Botev et~al\mbox{.}(2010)]%
        {botev2010kernel}
\bibfield{author}{\bibinfo{person}{Zdravko~I Botev}, \bibinfo{person}{Joseph~F
  Grotowski}, {and} \bibinfo{person}{Dirk~P Kroese}.}
  \bibinfo{year}{2010}\natexlab{}.
\newblock \showarticletitle{Kernel density estimation via diffusion}.
\newblock \bibinfo{journal}{\emph{The annals of Statistics}}
  \bibinfo{volume}{38}, \bibinfo{number}{5} (\bibinfo{year}{2010}),
  \bibinfo{pages}{2916--2957}.
\newblock


\bibitem[Chang et~al\mbox{.}(2018)]%
        {chang2018content}
\bibfield{author}{\bibinfo{person}{Buru Chang}, \bibinfo{person}{Yonggyu Park},
  \bibinfo{person}{Donghyeon Park}, \bibinfo{person}{Seongsoon Kim}, {and}
  \bibinfo{person}{Jaewoo Kang}.} \bibinfo{year}{2018}\natexlab{}.
\newblock \showarticletitle{Content-aware hierarchical point-of-interest
  embedding model for successive POI recommendation.}. In
  \bibinfo{booktitle}{\emph{IJCAI}}. \bibinfo{pages}{3301--3307}.
\newblock


\bibitem[Chen et~al\mbox{.}(2018)]%
        {chen2018privacy}
\bibfield{author}{\bibinfo{person}{Chaochao Chen}, \bibinfo{person}{Ziqi Liu},
  \bibinfo{person}{Peilin Zhao}, \bibinfo{person}{Jun Zhou}, {and}
  \bibinfo{person}{Xiaolong Li}.} \bibinfo{year}{2018}\natexlab{}.
\newblock \showarticletitle{Privacy preserving point-of-interest recommendation
  using decentralized matrix factorization}. In
  \bibinfo{booktitle}{\emph{Proceedings of the AAAI Conference on Artificial
  Intelligence}}, Vol.~\bibinfo{volume}{32}.
\newblock


\bibitem[Chen et~al\mbox{.}(2020)]%
        {chen2020multi}
\bibfield{author}{\bibinfo{person}{Hongxu Chen}, \bibinfo{person}{Hongzhi Yin},
  \bibinfo{person}{Xiangguo Sun}, \bibinfo{person}{Tong Chen},
  \bibinfo{person}{Bogdan Gabrys}, {and} \bibinfo{person}{Katarzyna Musial}.}
  \bibinfo{year}{2020}\natexlab{}.
\newblock \showarticletitle{Multi-level graph convolutional networks for
  cross-platform anchor link prediction}. In
  \bibinfo{booktitle}{\emph{Proceedings of the 26th ACM SIGKDD international
  conference on knowledge discovery \& data mining}}.
  \bibinfo{pages}{1503--1511}.
\newblock


\bibitem[Cheng et~al\mbox{.}(2013)]%
        {10.5555/2540128.2540504}
\bibfield{author}{\bibinfo{person}{Chen Cheng}, \bibinfo{person}{Haiqin Yang},
  \bibinfo{person}{Michael~R. Lyu}, {and} \bibinfo{person}{Irwin King}.}
  \bibinfo{year}{2013}\natexlab{}.
\newblock \showarticletitle{Where You like to Go next: Successive
  Point-of-Interest Recommendation}. In \bibinfo{booktitle}{\emph{Proceedings
  of the Twenty-Third International Joint Conference on Artificial
  Intelligence}} (Beijing, China) \emph{(\bibinfo{series}{IJCAI '13})}.
  \bibinfo{publisher}{AAAI Press}, \bibinfo{pages}{2605–2611}.
\newblock
\showISBNx{9781577356332}


\bibitem[Davies et~al\mbox{.}(1981)]%
        {davies1981constituents}
\bibfield{author}{\bibinfo{person}{Jack~N Davies}, \bibinfo{person}{Graeme~E
  Hobson}, {and} \bibinfo{person}{WB McGlasson}.}
  \bibinfo{year}{1981}\natexlab{}.
\newblock \showarticletitle{The constituents of tomato fruit—the influence of
  environment, nutrition, and genotype}.
\newblock \bibinfo{journal}{\emph{Critical Reviews in Food Science \&
  Nutrition}} \bibinfo{volume}{15}, \bibinfo{number}{3} (\bibinfo{year}{1981}),
  \bibinfo{pages}{205--280}.
\newblock


\bibitem[Duriakova et~al\mbox{.}(2019)]%
        {duriakova2019pdmfrec}
\bibfield{author}{\bibinfo{person}{Erika Duriakova}, \bibinfo{person}{Elias~Z
  Tragos}, \bibinfo{person}{Barry Smyth}, \bibinfo{person}{Neil Hurley},
  \bibinfo{person}{Francisco~J Pe{\~n}a}, \bibinfo{person}{Panagiotis
  Symeonidis}, \bibinfo{person}{James Geraci}, {and} \bibinfo{person}{Aonghus
  Lawlor}.} \bibinfo{year}{2019}\natexlab{}.
\newblock \showarticletitle{PDMFRec: a decentralised matrix factorisation with
  tunable user-centric privacy}. In \bibinfo{booktitle}{\emph{Proceedings of
  the 13th ACM Conference on Recommender Systems}}. \bibinfo{pages}{457--461}.
\newblock


\bibitem[Dwork et~al\mbox{.}(2014)]%
        {dwork2014algorithmic}
\bibfield{author}{\bibinfo{person}{Cynthia Dwork}, \bibinfo{person}{Aaron
  Roth}, {et~al\mbox{.}}} \bibinfo{year}{2014}\natexlab{}.
\newblock \showarticletitle{The algorithmic foundations of differential
  privacy.}
\newblock \bibinfo{journal}{\emph{Found. Trends Theor. Comput. Sci.}}
  \bibinfo{volume}{9}, \bibinfo{number}{3-4} (\bibinfo{year}{2014}),
  \bibinfo{pages}{211--407}.
\newblock


\bibitem[Fan and Khademi(2014)]%
        {DBLP:journals/corr/FanK14}
\bibfield{author}{\bibinfo{person}{Mingming Fan} {and} \bibinfo{person}{Maryam
  Khademi}.} \bibinfo{year}{2014}\natexlab{}.
\newblock \showarticletitle{Predicting a Business Star in Yelp from Its Reviews
  Text Alone}.
\newblock \bibinfo{journal}{\emph{CoRR}}  \bibinfo{volume}{abs/1401.0864}
  (\bibinfo{year}{2014}).
\newblock
\showeprint[arXiv]{1401.0864}
\urldef\tempurl%
\url{http://arxiv.org/abs/1401.0864}
\showURL{%
\tempurl}


\bibitem[Feng et~al\mbox{.}(2018a)]%
        {FengJie2018DPhm}
\bibfield{author}{\bibinfo{person}{Jie Feng}, \bibinfo{person}{Yong Li},
  \bibinfo{person}{Chao Zhang}, \bibinfo{person}{Funing Sun},
  \bibinfo{person}{Fanchao Meng}, \bibinfo{person}{Ang Guo}, {and}
  \bibinfo{person}{Depeng Jin}.} \bibinfo{year}{2018}\natexlab{a}.
\newblock \showarticletitle{DeepMove: Predicting human mobility with
  attentional recurrent networks}. In \bibinfo{booktitle}{\emph{The Web
  Conference 2018 - Proceedings of the World Wide Web Conference, WWW 2018}}.
  \bibinfo{pages}{1459--1468}.
\newblock
\showISBNx{1450356397}


\bibitem[Feng et~al\mbox{.}(2018b)]%
        {10.1145/3178876.3186058}
\bibfield{author}{\bibinfo{person}{Jie Feng}, \bibinfo{person}{Yong Li},
  \bibinfo{person}{Chao Zhang}, \bibinfo{person}{Funing Sun},
  \bibinfo{person}{Fanchao Meng}, \bibinfo{person}{Ang Guo}, {and}
  \bibinfo{person}{Depeng Jin}.} \bibinfo{year}{2018}\natexlab{b}.
\newblock \showarticletitle{DeepMove: Predicting Human Mobility with
  Attentional Recurrent Networks}. In \bibinfo{booktitle}{\emph{Proceedings of
  the 2018 World Wide Web Conference}} (Lyon, France)
  \emph{(\bibinfo{series}{WWW '18})}. \bibinfo{publisher}{International World
  Wide Web Conferences Steering Committee}, \bibinfo{address}{Republic and
  Canton of Geneva, CHE}, \bibinfo{pages}{1459–1468}.
\newblock
\showISBNx{9781450356398}
\urldef\tempurl%
\url{https://doi.org/10.1145/3178876.3186058}
\showDOI{\tempurl}


\bibitem[Giantomassi et~al\mbox{.}(2015)]%
        {Giantomassi2015ElectricMF}
\bibfield{author}{\bibinfo{person}{Andrea Giantomassi},
  \bibinfo{person}{Francesco Ferracuti}, \bibinfo{person}{Sabrina Iarlori},
  \bibinfo{person}{Gianluca Ippoliti}, {and} \bibinfo{person}{Sauro Longhi}.}
  \bibinfo{year}{2015}\natexlab{}.
\newblock \showarticletitle{Electric Motor Fault Detection and Diagnosis by
  Kernel Density Estimation and Kullback-Leibler Divergence Based on Stator
  Current Measurements}.
\newblock \bibinfo{journal}{\emph{IEEE Transactions on Industrial Electronics}}
   \bibinfo{volume}{62} (\bibinfo{year}{2015}), \bibinfo{pages}{1770--1780}.
\newblock


\bibitem[Guo et~al\mbox{.}(2020)]%
        {guo2020attentional}
\bibfield{author}{\bibinfo{person}{Qing Guo}, \bibinfo{person}{Zhu Sun},
  \bibinfo{person}{Jie Zhang}, {and} \bibinfo{person}{Yin-Leng Theng}.}
  \bibinfo{year}{2020}\natexlab{}.
\newblock \showarticletitle{An attentional recurrent neural network for
  personalized next location recommendation}. In
  \bibinfo{booktitle}{\emph{Proceedings of the AAAI Conference on artificial
  intelligence}}, Vol.~\bibinfo{volume}{34}. \bibinfo{pages}{83--90}.
\newblock


\bibitem[Guo et~al\mbox{.}(2021)]%
        {GuoYeting2021PPRw}
\bibfield{author}{\bibinfo{person}{Yeting Guo}, \bibinfo{person}{Fang Liu},
  \bibinfo{person}{Zhiping Cai}, \bibinfo{person}{Hui Zeng},
  \bibinfo{person}{Li Chen}, \bibinfo{person}{Tongqing Zhou}, {and}
  \bibinfo{person}{Nong Xiao}.} \bibinfo{year}{2021}\natexlab{}.
\newblock \showarticletitle{PREFER: Point-of-interest REcommendation with
  efficiency and privacy-preservation via Federated Edge leaRning}.
\newblock \bibinfo{journal}{\emph{Proceedings of ACM on interactive, mobile,
  wearable and ubiquitous technologies}} \bibinfo{volume}{5},
  \bibinfo{number}{1} (\bibinfo{year}{2021}), \bibinfo{pages}{1--25}.
\newblock
\showISSN{2474-9567}


\bibitem[Han et~al\mbox{.}(2020)]%
        {han2020contextualized}
\bibfield{author}{\bibinfo{person}{Peng Han}, \bibinfo{person}{Zhongxiao Li},
  \bibinfo{person}{Yong Liu}, \bibinfo{person}{Peilin Zhao},
  \bibinfo{person}{Jing Li}, \bibinfo{person}{Hao Wang}, {and}
  \bibinfo{person}{Shuo Shang}.} \bibinfo{year}{2020}\natexlab{}.
\newblock \showarticletitle{Contextualized point-of-interest recommendation}.
  International Joint Conferences on Artificial Intelligence.
\newblock


\bibitem[Hochreiter and Schmidhuber(1997)]%
        {hochreiter1997long}
\bibfield{author}{\bibinfo{person}{Sepp Hochreiter} {and}
  \bibinfo{person}{J{\"u}rgen Schmidhuber}.} \bibinfo{year}{1997}\natexlab{}.
\newblock \showarticletitle{Long short-term memory}.
\newblock \bibinfo{journal}{\emph{Neural computation}} \bibinfo{volume}{9},
  \bibinfo{number}{8} (\bibinfo{year}{1997}), \bibinfo{pages}{1735--1780}.
\newblock


\bibitem[Hull et~al\mbox{.}(2010)]%
        {HullGordon2010Cgpi}
\bibfield{author}{\bibinfo{person}{Gordon Hull},
  \bibinfo{person}{Heather~Richter Lipford}, {and} \bibinfo{person}{Celine
  Latulipe}.} \bibinfo{year}{2010}\natexlab{}.
\newblock \showarticletitle{Contextual gaps: privacy issues on Facebook}.
\newblock \bibinfo{journal}{\emph{Ethics and information technology}}
  \bibinfo{volume}{13}, \bibinfo{number}{4} (\bibinfo{year}{2010}),
  \bibinfo{pages}{289--302}.
\newblock
\showISSN{1388-1957}


\bibitem[Kalathil et~al\mbox{.}(2014)]%
        {kalathil2014decentralized}
\bibfield{author}{\bibinfo{person}{Dileep Kalathil}, \bibinfo{person}{Naumaan
  Nayyar}, {and} \bibinfo{person}{Rahul Jain}.}
  \bibinfo{year}{2014}\natexlab{}.
\newblock \showarticletitle{Decentralized learning for multiplayer multiarmed
  bandits}.
\newblock \bibinfo{journal}{\emph{IEEE Transactions on Information Theory}}
  \bibinfo{volume}{60}, \bibinfo{number}{4} (\bibinfo{year}{2014}),
  \bibinfo{pages}{2331--2345}.
\newblock


\bibitem[Kang and McAuley(2018)]%
        {kang2018self}
\bibfield{author}{\bibinfo{person}{Wang-Cheng Kang} {and}
  \bibinfo{person}{Julian McAuley}.} \bibinfo{year}{2018}\natexlab{}.
\newblock \showarticletitle{Self-attentive sequential recommendation}. In
  \bibinfo{booktitle}{\emph{2018 IEEE International Conference on Data Mining
  (ICDM)}}. IEEE, \bibinfo{pages}{197--206}.
\newblock


\bibitem[Kong et~al\mbox{.}(2019)]%
        {kong2019mutual}
\bibfield{author}{\bibinfo{person}{Lingpeng Kong}, \bibinfo{person}{Cyprien
  de~Masson d'Autume}, \bibinfo{person}{Wang Ling}, \bibinfo{person}{Lei Yu},
  \bibinfo{person}{Zihang Dai}, {and} \bibinfo{person}{Dani Yogatama}.}
  \bibinfo{year}{2019}\natexlab{}.
\newblock \showarticletitle{A mutual information maximization perspective of
  language representation learning}.
\newblock \bibinfo{journal}{\emph{arXiv preprint arXiv:1910.08350}}
  (\bibinfo{year}{2019}).
\newblock


\bibitem[Krichene and Rendle(2020)]%
        {krichene2020sampled}
\bibfield{author}{\bibinfo{person}{Walid Krichene} {and}
  \bibinfo{person}{Steffen Rendle}.} \bibinfo{year}{2020}\natexlab{}.
\newblock \showarticletitle{On sampled metrics for item recommendation}. In
  \bibinfo{booktitle}{\emph{Proceedings of the 26th ACM SIGKDD international
  conference on knowledge discovery \& data mining}}.
  \bibinfo{pages}{1748--1757}.
\newblock


\bibitem[Leng et~al\mbox{.}(2015)]%
        {leng2015hashing}
\bibfield{author}{\bibinfo{person}{Cong Leng}, \bibinfo{person}{Jiaxiang Wu},
  \bibinfo{person}{Jian Cheng}, \bibinfo{person}{Xi Zhang}, {and}
  \bibinfo{person}{Hanqing Lu}.} \bibinfo{year}{2015}\natexlab{}.
\newblock \showarticletitle{Hashing for distributed data}. In
  \bibinfo{booktitle}{\emph{International Conference on Machine Learning}}.
  PMLR, \bibinfo{pages}{1642--1650}.
\newblock


\bibitem[Li et~al\mbox{.}(2018)]%
        {Li2018NextPR}
\bibfield{author}{\bibinfo{person}{Ranzhen Li}, \bibinfo{person}{Yanyan Shen},
  {and} \bibinfo{person}{Yanmin Zhu}.} \bibinfo{year}{2018}\natexlab{}.
\newblock \showarticletitle{Next Point-of-Interest Recommendation with Temporal
  and Multi-level Context Attention}.
\newblock \bibinfo{journal}{\emph{2018 IEEE International Conference on Data
  Mining (ICDM)}} (\bibinfo{year}{2018}), \bibinfo{pages}{1110--1115}.
\newblock


\bibitem[Li et~al\mbox{.}(2021)]%
        {LiYang2021DCSf}
\bibfield{author}{\bibinfo{person}{Yang Li}, \bibinfo{person}{Tong Chen},
  \bibinfo{person}{Hongzhi Yin}, {and} \bibinfo{person}{Zi Huang}.}
  \bibinfo{year}{2021}\natexlab{}.
\newblock \showarticletitle{Discovering Collaborative Signals for Next POI
  Recommendation with Iterative Seq2Graph Augmentation}.
\newblock  (\bibinfo{year}{2021}).
\newblock


\bibitem[Lian et~al\mbox{.}(2020a)]%
        {10.1145/3394486.3403252}
\bibfield{author}{\bibinfo{person}{Defu Lian}, \bibinfo{person}{Yongji Wu},
  \bibinfo{person}{Yong Ge}, \bibinfo{person}{Xing Xie}, {and}
  \bibinfo{person}{Enhong Chen}.} \bibinfo{year}{2020}\natexlab{a}.
\newblock \bibinfo{booktitle}{\emph{Geography-Aware Sequential Location
  Recommendation}}.
\newblock \bibinfo{publisher}{Association for Computing Machinery},
  \bibinfo{address}{New York, NY, USA}, \bibinfo{pages}{2009–2019}.
\newblock
\showISBNx{9781450379984}
\urldef\tempurl%
\url{https://doi.org/10.1145/3394486.3403252}
\showURL{%
\tempurl}


\bibitem[Lian et~al\mbox{.}(2020b)]%
        {lian2020geography}
\bibfield{author}{\bibinfo{person}{Defu Lian}, \bibinfo{person}{Yongji Wu},
  \bibinfo{person}{Yong Ge}, \bibinfo{person}{Xing Xie}, {and}
  \bibinfo{person}{Enhong Chen}.} \bibinfo{year}{2020}\natexlab{b}.
\newblock \showarticletitle{Geography-aware sequential location
  recommendation}. In \bibinfo{booktitle}{\emph{Proceedings of the 26th ACM
  SIGKDD international conference on knowledge discovery \& data mining}}.
  \bibinfo{pages}{2009--2019}.
\newblock


\bibitem[Lian et~al\mbox{.}(2014)]%
        {10.1145/2623330.2623638}
\bibfield{author}{\bibinfo{person}{Defu Lian}, \bibinfo{person}{Cong Zhao},
  \bibinfo{person}{Xing Xie}, \bibinfo{person}{Guangzhong Sun},
  \bibinfo{person}{Enhong Chen}, {and} \bibinfo{person}{Yong Rui}.}
  \bibinfo{year}{2014}\natexlab{}.
\newblock \showarticletitle{GeoMF: Joint Geographical Modeling and Matrix
  Factorization for Point-of-Interest Recommendation}. In
  \bibinfo{booktitle}{\emph{Proceedings of the 20th ACM SIGKDD International
  Conference on Knowledge Discovery and Data Mining}} (New York, New York, USA)
  \emph{(\bibinfo{series}{KDD '14})}. \bibinfo{publisher}{Association for
  Computing Machinery}, \bibinfo{address}{New York, NY, USA},
  \bibinfo{pages}{831–840}.
\newblock
\showISBNx{9781450329569}
\urldef\tempurl%
\url{https://doi.org/10.1145/2623330.2623638}
\showDOI{\tempurl}


\bibitem[Liao et~al\mbox{.}(2010)]%
        {liao2010network}
\bibfield{author}{\bibinfo{person}{Yongjun Liao}, \bibinfo{person}{Pierre
  Geurts}, {and} \bibinfo{person}{Guy Leduc}.} \bibinfo{year}{2010}\natexlab{}.
\newblock \showarticletitle{Network distance prediction based on decentralized
  matrix factorization}. In \bibinfo{booktitle}{\emph{International Conference
  on Research in Networking}}. Springer, \bibinfo{pages}{15--26}.
\newblock


\bibitem[Liu et~al\mbox{.}(2016)]%
        {10.5555/3015812.3015841}
\bibfield{author}{\bibinfo{person}{Qiang Liu}, \bibinfo{person}{Shu Wu},
  \bibinfo{person}{Liang Wang}, {and} \bibinfo{person}{Tieniu Tan}.}
  \bibinfo{year}{2016}\natexlab{}.
\newblock \showarticletitle{Predicting the next Location: A Recurrent Model
  with Spatial and Temporal Contexts}. In \bibinfo{booktitle}{\emph{Proceedings
  of the Thirtieth AAAI Conference on Artificial Intelligence}} (Phoenix,
  Arizona) \emph{(\bibinfo{series}{AAAI'16})}. \bibinfo{publisher}{AAAI Press},
  \bibinfo{pages}{194–200}.
\newblock


\bibitem[Liu et~al\mbox{.}(2013)]%
        {liu2013personalized}
\bibfield{author}{\bibinfo{person}{Xin Liu}, \bibinfo{person}{Yong Liu},
  \bibinfo{person}{Karl Aberer}, {and} \bibinfo{person}{Chunyan Miao}.}
  \bibinfo{year}{2013}\natexlab{}.
\newblock \showarticletitle{Personalized point-of-interest recommendation by
  mining users' preference transition}. In
  \bibinfo{booktitle}{\emph{Proceedings of the 22nd ACM international
  conference on Information \& Knowledge Management}}.
  \bibinfo{pages}{733--738}.
\newblock


\bibitem[Liu et~al\mbox{.}(2014)]%
        {liu2014exploiting}
\bibfield{author}{\bibinfo{person}{Yong Liu}, \bibinfo{person}{Wei Wei},
  \bibinfo{person}{Aixin Sun}, {and} \bibinfo{person}{Chunyan Miao}.}
  \bibinfo{year}{2014}\natexlab{}.
\newblock \showarticletitle{Exploiting geographical neighborhood
  characteristics for location recommendation}. In
  \bibinfo{booktitle}{\emph{Proceedings of the 23rd ACM international
  conference on conference on information and knowledge management}}.
  \bibinfo{pages}{739--748}.
\newblock


\bibitem[Long et~al\mbox{.}(2017)]%
        {LongYan2017SPRE}
\bibfield{author}{\bibinfo{person}{Yan Long}, \bibinfo{person}{Pengpeng Zhao},
  \bibinfo{person}{Victor~S Sheng}, \bibinfo{person}{Guanfeng Liu},
  \bibinfo{person}{Jiajie Xu}, \bibinfo{person}{Jian Wu}, {and}
  \bibinfo{person}{Zhiming Cui}.} \bibinfo{year}{2017}\natexlab{}.
\newblock \showarticletitle{Social Personalized Ranking Embedding for Next POI
  Recommendation}.
\newblock In \bibinfo{booktitle}{\emph{Web Information Systems Engineering -
  WISE 2017}}. \bibinfo{series}{Lecture Notes in Computer Science},
  Vol.~\bibinfo{volume}{10569}. \bibinfo{publisher}{Springer International
  Publishing}, \bibinfo{address}{Cham}, \bibinfo{pages}{91--105}.
\newblock
\showISBNx{3319687824}
\showISSN{0302-9743}


\bibitem[Luo et~al\mbox{.}(2021a)]%
        {LuoYingtao2021SSan}
\bibfield{author}{\bibinfo{person}{Yingtao Luo}, \bibinfo{person}{Qiang Liu},
  {and} \bibinfo{person}{Zhaocheng Liu}.} \bibinfo{year}{2021}\natexlab{a}.
\newblock \showarticletitle{STAN: Spatio-temporal attention network for next
  location recommendation}. In \bibinfo{booktitle}{\emph{The Web Conference
  2021 - Proceedings of the World Wide Web Conference, WWW 2021}}.
  \bibinfo{pages}{2177--2185}.
\newblock
\showISBNx{9781450383127}


\bibitem[Luo et~al\mbox{.}(2021b)]%
        {luo2021stan}
\bibfield{author}{\bibinfo{person}{Yingtao Luo}, \bibinfo{person}{Qiang Liu},
  {and} \bibinfo{person}{Zhaocheng Liu}.} \bibinfo{year}{2021}\natexlab{b}.
\newblock \showarticletitle{Stan: Spatio-temporal attention network for next
  location recommendation}. In \bibinfo{booktitle}{\emph{Proceedings of the Web
  Conference 2021}}. \bibinfo{pages}{2177--2185}.
\newblock


\bibitem[MacQueen et~al\mbox{.}(1967)]%
        {macqueen1967some}
\bibfield{author}{\bibinfo{person}{James MacQueen} {et~al\mbox{.}}}
  \bibinfo{year}{1967}\natexlab{}.
\newblock \showarticletitle{Some methods for classification and analysis of
  multivariate observations}. In \bibinfo{booktitle}{\emph{Proceedings of the
  fifth Berkeley symposium on mathematical statistics and probability}},
  Vol.~\bibinfo{volume}{1}. Oakland, CA, USA, \bibinfo{pages}{281--297}.
\newblock


\bibitem[Narayanan and Shmatikov(2008)]%
        {NarayananA2008RDoL}
\bibfield{author}{\bibinfo{person}{A Narayanan} {and} \bibinfo{person}{V
  Shmatikov}.} \bibinfo{year}{2008}\natexlab{}.
\newblock \showarticletitle{Robust De-anonymization of Large Sparse Datasets}.
  In \bibinfo{booktitle}{\emph{2008 IEEE Symposium on Security and Privacy (sp
  2008)}}. \bibinfo{publisher}{IEEE}, \bibinfo{pages}{111--125}.
\newblock
\showISBNx{9780769531687}
\showISSN{1081-6011}


\bibitem[Nguyen et~al\mbox{.}(2017a)]%
        {nguyen2017argument}
\bibfield{author}{\bibinfo{person}{Quoc Viet~Hung Nguyen},
  \bibinfo{person}{Chi~Thang Duong}, \bibinfo{person}{Thanh~Tam Nguyen},
  \bibinfo{person}{Matthias Weidlich}, \bibinfo{person}{Karl Aberer},
  \bibinfo{person}{Hongzhi Yin}, {and} \bibinfo{person}{Xiaofang Zhou}.}
  \bibinfo{year}{2017}\natexlab{a}.
\newblock \showarticletitle{Argument discovery via crowdsourcing}.
\newblock \bibinfo{journal}{\emph{The VLDB Journal}} \bibinfo{volume}{26},
  \bibinfo{number}{4} (\bibinfo{year}{2017}), \bibinfo{pages}{511--535}.
\newblock


\bibitem[Nguyen et~al\mbox{.}(2018)]%
        {8509453}
\bibfield{author}{\bibinfo{person}{Quoc Viet~Hung Nguyen},
  \bibinfo{person}{Huu~Viet Huynh}, \bibinfo{person}{Thanh~Tam Nguyen},
  \bibinfo{person}{Matthias Weidlich}, \bibinfo{person}{Hongzhi Yin}, {and}
  \bibinfo{person}{Xiaofang Zhou}.} \bibinfo{year}{2018}\natexlab{}.
\newblock \showarticletitle{Computing Crowd Consensus with Partial Agreement}.
  In \bibinfo{booktitle}{\emph{2018 IEEE 34th International Conference on Data
  Engineering (ICDE)}}. \bibinfo{pages}{1749--1750}.
\newblock
\urldef\tempurl%
\url{https://doi.org/10.1109/ICDE.2018.00232}
\showDOI{\tempurl}


\bibitem[Nguyen et~al\mbox{.}(2017b)]%
        {nguyen2017retaining}
\bibfield{author}{\bibinfo{person}{Thanh~Tam Nguyen},
  \bibinfo{person}{Chi~Thang Duong}, \bibinfo{person}{Matthias Weidlich},
  \bibinfo{person}{Hongzhi Yin}, {and} \bibinfo{person}{Quoc Viet~Hung
  Nguyen}.} \bibinfo{year}{2017}\natexlab{b}.
\newblock \showarticletitle{Retaining data from streams of social platforms
  with minimal regret}. In \bibinfo{booktitle}{\emph{Twenty-sixth International
  Joint Conference on Artificial Intelligence}}.
\newblock


\bibitem[Rao et~al\mbox{.}(2021)]%
        {rao2021privacy}
\bibfield{author}{\bibinfo{person}{Jinmeng Rao}, \bibinfo{person}{Song Gao},
  \bibinfo{person}{Mingxiao Li}, {and} \bibinfo{person}{Qunying Huang}.}
  \bibinfo{year}{2021}\natexlab{}.
\newblock \showarticletitle{A privacy-preserving framework for location
  recommendation using decentralized collaborative machine learning}.
\newblock \bibinfo{journal}{\emph{Transactions in GIS}} \bibinfo{volume}{25},
  \bibinfo{number}{3} (\bibinfo{year}{2021}), \bibinfo{pages}{1153--1175}.
\newblock


\bibitem[Rendle(2012)]%
        {RendleSteffen2012FMwl}
\bibfield{author}{\bibinfo{person}{Steffen Rendle}.}
  \bibinfo{year}{2012}\natexlab{}.
\newblock \showarticletitle{Factorization Machines with libFM}.
\newblock \bibinfo{journal}{\emph{ACM transactions on intelligent systems and
  technology}} \bibinfo{volume}{3}, \bibinfo{number}{3} (\bibinfo{year}{2012}),
  \bibinfo{pages}{1--22}.
\newblock
\showISSN{2157-6904}


\bibitem[Robusto(1957)]%
        {Robusto1957TheCF}
\bibfield{author}{\bibinfo{person}{C Robusto}.}
  \bibinfo{year}{1957}\natexlab{}.
\newblock \showarticletitle{The Cosine-Haversine Formula}.
\newblock \bibinfo{journal}{\emph{Amer. Math. Monthly}}  \bibinfo{volume}{64}
  (\bibinfo{year}{1957}), \bibinfo{pages}{38}.
\newblock


\bibitem[Wang et~al\mbox{.}(2020b)]%
        {WangQinyong2020NPRo}
\bibfield{author}{\bibinfo{person}{Qinyong Wang}, \bibinfo{person}{Hongzhi
  Yin}, \bibinfo{person}{Tong Chen}, \bibinfo{person}{Zi Huang},
  \bibinfo{person}{Hao Wang}, \bibinfo{person}{Yanchang Zhao}, {and}
  \bibinfo{person}{Nguyen~Quoc Viet~Hung}.} \bibinfo{year}{2020}\natexlab{b}.
\newblock \showarticletitle{Next Point-of-Interest Recommendation on
  Resource-Constrained Mobile Devices}. In
  \bibinfo{booktitle}{\emph{Proceedings of The Web Conference 2020}}
  \emph{(\bibinfo{series}{WWW '20})}. \bibinfo{publisher}{ACM},
  \bibinfo{pages}{906--916}.
\newblock
\showISBNx{1450370233}


\bibitem[Wang et~al\mbox{.}(2021)]%
        {wang2021fast}
\bibfield{author}{\bibinfo{person}{Qinyong Wang}, \bibinfo{person}{Hongzhi
  Yin}, \bibinfo{person}{Tong Chen}, \bibinfo{person}{Junliang Yu},
  \bibinfo{person}{Alexander Zhou}, {and} \bibinfo{person}{Xiangliang Zhang}.}
  \bibinfo{year}{2021}\natexlab{}.
\newblock \showarticletitle{Fast-adapting and privacy-preserving federated
  recommender system}.
\newblock \bibinfo{journal}{\emph{The VLDB Journal}} (\bibinfo{year}{2021}),
  \bibinfo{pages}{1--20}.
\newblock


\bibitem[Wang et~al\mbox{.}(2018)]%
        {wang2018neural}
\bibfield{author}{\bibinfo{person}{Qinyong Wang}, \bibinfo{person}{Hongzhi
  Yin}, \bibinfo{person}{Zhiting Hu}, \bibinfo{person}{Defu Lian},
  \bibinfo{person}{Hao Wang}, {and} \bibinfo{person}{Zi Huang}.}
  \bibinfo{year}{2018}\natexlab{}.
\newblock \showarticletitle{Neural memory streaming recommender networks with
  adversarial training}. In \bibinfo{booktitle}{\emph{Proceedings of the 24th
  ACM SIGKDD International Conference on Knowledge Discovery \& Data Mining}}.
  \bibinfo{pages}{2467--2475}.
\newblock


\bibitem[Wang et~al\mbox{.}(2019)]%
        {wang2019enhancing}
\bibfield{author}{\bibinfo{person}{Qinyong Wang}, \bibinfo{person}{Hongzhi
  Yin}, \bibinfo{person}{Hao Wang}, \bibinfo{person}{Quoc Viet~Hung Nguyen},
  \bibinfo{person}{Zi Huang}, {and} \bibinfo{person}{Lizhen Cui}.}
  \bibinfo{year}{2019}\natexlab{}.
\newblock \showarticletitle{Enhancing collaborative filtering with generative
  augmentation}. In \bibinfo{booktitle}{\emph{Proceedings of the 25th ACM
  SIGKDD International Conference on Knowledge Discovery \& Data Mining}}.
  \bibinfo{pages}{548--556}.
\newblock


\bibitem[Wang et~al\mbox{.}(2020a)]%
        {wang2020group}
\bibfield{author}{\bibinfo{person}{Xiwei Wang}, \bibinfo{person}{Minh Nguyen},
  \bibinfo{person}{Jonathan Carr}, \bibinfo{person}{Longyin Cui}, {and}
  \bibinfo{person}{Kiho Lim}.} \bibinfo{year}{2020}\natexlab{a}.
\newblock \showarticletitle{A group preference-based privacy-preserving POI
  recommender system}.
\newblock \bibinfo{journal}{\emph{ICT Express}} \bibinfo{volume}{6},
  \bibinfo{number}{3} (\bibinfo{year}{2020}), \bibinfo{pages}{204--208}.
\newblock


\bibitem[Wei et~al\mbox{.}(2020)]%
        {wei2020federated}
\bibfield{author}{\bibinfo{person}{Kang Wei}, \bibinfo{person}{Jun Li},
  \bibinfo{person}{Ming Ding}, \bibinfo{person}{Chuan Ma},
  \bibinfo{person}{Howard~H Yang}, \bibinfo{person}{Farhad Farokhi},
  \bibinfo{person}{Shi Jin}, \bibinfo{person}{Tony~QS Quek}, {and}
  \bibinfo{person}{H~Vincent Poor}.} \bibinfo{year}{2020}\natexlab{}.
\newblock \showarticletitle{Federated learning with differential privacy:
  Algorithms and performance analysis}.
\newblock \bibinfo{journal}{\emph{IEEE Transactions on Information Forensics
  and Security}}  \bibinfo{volume}{15} (\bibinfo{year}{2020}),
  \bibinfo{pages}{3454--3469}.
\newblock


\bibitem[Weimer et~al\mbox{.}(2007)]%
        {weimer2007cofi}
\bibfield{author}{\bibinfo{person}{Markus Weimer}, \bibinfo{person}{Alexandros
  Karatzoglou}, \bibinfo{person}{Quoc Le}, {and} \bibinfo{person}{Alex Smola}.}
  \bibinfo{year}{2007}\natexlab{}.
\newblock \showarticletitle{Cofi rank-maximum margin matrix factorization for
  collaborative ranking}.
\newblock \bibinfo{journal}{\emph{Advances in neural information processing
  systems}}  \bibinfo{volume}{20} (\bibinfo{year}{2007}).
\newblock


\bibitem[Yao et~al\mbox{.}(2017)]%
        {yao2017serm}
\bibfield{author}{\bibinfo{person}{Di Yao}, \bibinfo{person}{Chao Zhang},
  \bibinfo{person}{Jianhui Huang}, {and} \bibinfo{person}{Jingping Bi}.}
  \bibinfo{year}{2017}\natexlab{}.
\newblock \showarticletitle{Serm: A recurrent model for next location
  prediction in semantic trajectories}. In
  \bibinfo{booktitle}{\emph{Proceedings of the 2017 ACM on Conference on
  Information and Knowledge Management}}. \bibinfo{pages}{2411--2414}.
\newblock


\bibitem[Ye et~al\mbox{.}(2021)]%
        {ye2021personalized}
\bibfield{author}{\bibinfo{person}{Guanhua Ye}, \bibinfo{person}{Hongzhi Yin},
  \bibinfo{person}{Tong Chen}, \bibinfo{person}{Miao Xu}, \bibinfo{person}{Quoc
  Viet~Hung Nguyen}, {and} \bibinfo{person}{Jiangning Song}.}
  \bibinfo{year}{2021}\natexlab{}.
\newblock \bibinfo{title}{Personalized On-Device E-health Analytics with
  Decentralized Block Coordinate Descent}.
\newblock
\newblock
\showeprint[arxiv]{2112.09341}~[cs.LG]


\bibitem[Ye et~al\mbox{.}(2022)]%
        {ye2022personalized}
\bibfield{author}{\bibinfo{person}{Guanhua Ye}, \bibinfo{person}{Hongzhi Yin},
  \bibinfo{person}{Tong Chen}, \bibinfo{person}{Miao Xu}, \bibinfo{person}{Quoc
  Viet~Hung Nguyen}, {and} \bibinfo{person}{Jiangning Song}.}
  \bibinfo{year}{2022}\natexlab{}.
\newblock \showarticletitle{Personalized on-device e-health analytics with
  decentralized block coordinate descent}.
\newblock \bibinfo{journal}{\emph{IEEE Journal of Biomedical and Health
  Informatics}} (\bibinfo{year}{2022}).
\newblock


\bibitem[Yu et~al\mbox{.}(2019)]%
        {yu2019generating}
\bibfield{author}{\bibinfo{person}{Junliang Yu}, \bibinfo{person}{Min Gao},
  \bibinfo{person}{Hongzhi Yin}, \bibinfo{person}{Jundong Li},
  \bibinfo{person}{Chongming Gao}, {and} \bibinfo{person}{Qinyong Wang}.}
  \bibinfo{year}{2019}\natexlab{}.
\newblock \showarticletitle{Generating reliable friends via adversarial
  training to improve social recommendation}. In \bibinfo{booktitle}{\emph{2019
  IEEE International Conference on Data Mining (ICDM)}}. IEEE,
  \bibinfo{pages}{768--777}.
\newblock


\bibitem[Yu et~al\mbox{.}(2020)]%
        {yu2020enhance}
\bibfield{author}{\bibinfo{person}{Junliang Yu}, \bibinfo{person}{Hongzhi Yin},
  \bibinfo{person}{Jundong Li}, \bibinfo{person}{Min Gao}, \bibinfo{person}{Zi
  Huang}, {and} \bibinfo{person}{Lizhen Cui}.} \bibinfo{year}{2020}\natexlab{}.
\newblock \showarticletitle{Enhance social recommendation with adversarial
  graph convolutional networks}.
\newblock \bibinfo{journal}{\emph{IEEE Transactions on Knowledge and Data
  Engineering}} (\bibinfo{year}{2020}).
\newblock


\bibitem[Zhang et~al\mbox{.}(2021)]%
        {zhang2021survey}
\bibfield{author}{\bibinfo{person}{Chen Zhang}, \bibinfo{person}{Yu Xie},
  \bibinfo{person}{Hang Bai}, \bibinfo{person}{Bin Yu},
  \bibinfo{person}{Weihong Li}, {and} \bibinfo{person}{Yuan Gao}.}
  \bibinfo{year}{2021}\natexlab{}.
\newblock \showarticletitle{A survey on federated learning}.
\newblock \bibinfo{journal}{\emph{Knowledge-Based Systems}}
  \bibinfo{volume}{216} (\bibinfo{year}{2021}), \bibinfo{pages}{106775}.
\newblock


\bibitem[Zhao et~al\mbox{.}(2020b)]%
        {zhao2020discovering}
\bibfield{author}{\bibinfo{person}{Kangzhi Zhao}, \bibinfo{person}{Yong Zhang},
  \bibinfo{person}{Hongzhi Yin}, \bibinfo{person}{Jin Wang},
  \bibinfo{person}{Kai Zheng}, \bibinfo{person}{Xiaofang Zhou}, {and}
  \bibinfo{person}{Chunxiao Xing}.} \bibinfo{year}{2020}\natexlab{b}.
\newblock \showarticletitle{Discovering Subsequence Patterns for Next POI
  Recommendation.}. In \bibinfo{booktitle}{\emph{IJCAI}}.
  \bibinfo{pages}{3216--3222}.
\newblock


\bibitem[Zhao et~al\mbox{.}(2020a)]%
        {9133505}
\bibfield{author}{\bibinfo{person}{Pengpeng Zhao}, \bibinfo{person}{Anjing
  Luo}, \bibinfo{person}{Yanchi Liu}, \bibinfo{person}{Fuzhen Zhuang},
  \bibinfo{person}{Jiajie Xu}, \bibinfo{person}{Zhixu Li},
  \bibinfo{person}{Victor~S. Sheng}, {and} \bibinfo{person}{Xiaofang Zhou}.}
  \bibinfo{year}{2020}\natexlab{a}.
\newblock \showarticletitle{Where to Go Next: A Spatio-Temporal Gated Network
  for Next POI Recommendation}.
\newblock \bibinfo{journal}{\emph{IEEE Transactions on Knowledge and Data
  Engineering}} (\bibinfo{year}{2020}), \bibinfo{pages}{1--1}.
\newblock
\urldef\tempurl%
\url{https://doi.org/10.1109/TKDE.2020.3007194}
\showDOI{\tempurl}


\bibitem[Zhao et~al\mbox{.}(2018)]%
        {ZhaoShenglin2018SSLR}
\bibfield{author}{\bibinfo{person}{Shenglin Zhao}, \bibinfo{person}{Michael~R
  Lyu}, {and} \bibinfo{person}{Irwin King}.} \bibinfo{year}{2018}\natexlab{}.
\newblock \showarticletitle{STELLAR: Spatial-Temporal Latent Ranking Model for
  Successive POI Recommendation}.
\newblock In \bibinfo{booktitle}{\emph{Point-of-Interest Recommendation in
  Location-Based Social Networks}}. \bibinfo{publisher}{Springer Singapore},
  \bibinfo{address}{Singapore}, \bibinfo{pages}{79--94}.
\newblock
\showISBNx{9811313482}
\showISSN{2191-5768}


\bibitem[Zhou et~al\mbox{.}(2020)]%
        {zhou2020s3}
\bibfield{author}{\bibinfo{person}{Kun Zhou}, \bibinfo{person}{Hui Wang},
  \bibinfo{person}{Wayne~Xin Zhao}, \bibinfo{person}{Yutao Zhu},
  \bibinfo{person}{Sirui Wang}, \bibinfo{person}{Fuzheng Zhang},
  \bibinfo{person}{Zhongyuan Wang}, {and} \bibinfo{person}{Ji-Rong Wen}.}
  \bibinfo{year}{2020}\natexlab{}.
\newblock \showarticletitle{S3-rec: Self-supervised learning for sequential
  recommendation with mutual information maximization}. In
  \bibinfo{booktitle}{\emph{Proceedings of the 29th ACM International
  Conference on Information \& Knowledge Management}}.
  \bibinfo{pages}{1893--1902}.
\newblock


\end{thebibliography}

\appendix

\end{document}